\documentclass[structabstract]{aa}

\usepackage{natbib}
\usepackage{graphicx}
\usepackage{amsmath}
\usepackage{amsbsy}
\usepackage{txfonts}

\bibpunct{(}{)}{;}{a}{}{,}     

\newcommand{\micron}{$\mu$m}

\newcommand{\ten}[1]{$10^{#1}$}
\newcommand{\scit}[2]{$#1\times10^{#2}$}
\newcommand{\scim}[2]{#1\times10^{#2}}

\newcommand{\ps}{s$^{-1}$}
\newcommand{\pcs}{cm$^{-2}$}
\newcommand{\pcc}{cm$^{-3}$}
\newcommand{\eq}[1]{Eq.\ (\ref{eq:#1})}

\newcommand{\fig}[1]{Fig.\ \ref{fig:#1}}
\newcommand{\figg}[1]{Figure \ref{fig:#1}}
\newcommand{\tb}[1]{Table \ref{tb:#1}}
\newcommand{\rx}[1]{Reaction (\ref{eq:#1})}

\newcommand{\pz}{\phantom{0}}

\newcommand{\tacc}{t_{\rm acc}}
\newcommand{\cs}{c_{\rm s}}

\newcommand{\av}{A_{\rm V}}
\newcommand{\nh}{n_{\rm H}}
\newcommand{\td}{T_{\rm d}}
\newcommand{\mh}{H$_2$}
\newcommand{\mhm}{{\rm H}_2}
\newcommand{\hp}{H$^+$}
\newcommand{\hhp}{H$_2^+$}
\newcommand{\hhhp}{H$_3^+$}
\newcommand{\w}{H$_2$O}
\newcommand{\wm}{{\rm H}_2{\rm O}}
\newcommand{\oop}{O$_2^+$}
\newcommand{\ohp}{OH$^+$}
\newcommand{\ohhhp}{H$_3$O$^+$}
\newcommand{\mo}{O$_2$}
\newcommand{\cp}{C$^+$}
\newcommand{\cdo}{CO$_2$}
\newcommand{\hcop}{HCO$^+$}
\newcommand{\chh}{CH$_2$}
\newcommand{\chhh}{CH$_3$}
\newcommand{\mep}{CH$_3^+$}
\newcommand{\meh}{CH$_4$}
\newcommand{\mehhp}{CH$_5^+$}
\newcommand{\mn}{N$_2$}
\newcommand{\nhh}{NH$_2$}
\newcommand{\nnhp}{N$_2$H$^+$}
\newcommand{\amh}{NH$_3$}
\newcommand{\amhhp}{NH$_4^+$}
\newcommand{\fmh}{H$_2$CO}
\newcommand{\meoh}{CH$_3$OH}
\newcommand{\mf}{HCOOCH$_3$}
\newcommand{\hhs}{H$_2$S}

\defcitealias{visser09a}{Paper I}
\newcommand{\vddd}{\citetalias{visser09a}}

\begin{document}

\title{The chemical history of molecules in circumstellar disks. II. Gas-phase species}

\author{
R. Visser \inst{1,2}
 \and S.D. Doty \inst{3}
 \and E.F. van Dishoeck \inst{1,4}
}

\institute{
Leiden Observatory, Leiden University, P.O.\ Box 9513, 2300 RA Leiden, the Netherlands
\and
Department of Astronomy, University of Michigan, 500 Church Street, Ann Arbor, MI 48109-1042, USA\\
  \email{visserr@umich.edu}
\and
Department of Physics and Astronomy, Denison University, Granville, OH 43023, USA
\and
Max-Planck-Institut f\"ur extraterrestrische Physik, Giessenbachstrasse 1, 85748 Garching, Germany
}

\titlerunning{The chemical history of molecules in circumstellar disks. II.}

\date{Accepted \today}


\abstract
{The chemical composition of a molecular cloud changes dramatically as it collapses to form a low-mass protostar and circumstellar disk. Two-dimensional (2D) chemodynamical models are required to properly study this process.} 
{The goal of this work is to follow, for the first time, the chemical evolution in two dimensions all the way from a pre-stellar core into a circumstellar disk. Of special interest is the question whether the chemical composition of the disk is a result of chemical processing during the collapse phase, or whether it is determined by in situ processing after the disk has formed.} 
{Our model combines a semi-analytical method to get 2D axisymmetric density and velocity structures with detailed radiative transfer calculations to get temperature profiles and UV fluxes. Material is followed in from the core to the disk and a full gas-phase chemistry network -- including freeze-out onto and evaporation from cold dust grains -- is evolved along these trajectories. The abundances thus obtained are compared to the results from a static disk model and to observations of comets.} 
{The chemistry during the collapse phase is dominated by a few key processes, such as the evaporation of CO or the photodissociation of \w\@. Depending on the physical conditions encountered along specific trajectories, some of these processes are absent. At the end of the collapse phase, the disk can be divided into zones with different chemical histories. The disk is not in chemical equilibrium at the end of the collapse, so care must be taken when choosing the initial abundances for stand-alone disk chemistry models. Our model results imply that comets must be formed from material with different chemical histories: some of it is strongly processed, some of it remains pristine. Variations between individual comets are possible if they formed at different positions or different times in the solar nebula.} 
{} 

\keywords{astrochemistry -- stars: formation -- circumstellar matter -- planetary systems: protoplanetary disks -- molecular processes}

\maketitle


\section{Introduction}
\label{sec:intro}
The formation of a low-mass protostar out of a cold molecular cloud is accompanied by large-scale changes in the chemical composition of the constituent gas and dust. Pre-stellar cloud cores are cold ($\sim$10 K), moderately dense ($\sim$\ten{4}--\ten{6} \pcc), and irradiated only by the ambient interstellar radiation field \citep{shu87a,difrancesco07a,bergin07b}. The material heats up as the core collapses, and the inner few hundred AU flatten out to form a circumstellar disk with densities that are orders of magnitude higher than in the parent cloud \citep{dullemond07b}. Meanwhile, the protostar infuses the disk and remnant envelope with large fluxes of ultraviolet and X-ray photons. The chemical changes arising from these evolving physical conditions have been analysed with one-dimensional spherical models \citep[e.g.,][]{aikawa08a}. However, two-dimensional models are required to properly describe the formation of the circumstellar disk and to analyse the full chemical evolution from core to disk.

This paper is the second in a series of publications aiming to model the chemical evolution from pre-stellar cores to circumstellar disks in two dimensions. The first paper \citep[hereafter Paper I]{visser09a} contained a detailed description of our semi-analytical model and an analysis of the gas and ice abundances of carbon monoxide (CO) and water (\w). Most CO was found to evaporate during the infall phase and to freeze out again in those parts of the disk that are colder than $\sim$20 K\@. The higher binding energy of \w{} keeps it in solid form at all times, except within $\sim$10 AU of the star. Based on the time that the infalling material spends at dust temperatures between 20 and 40 K, first-generation complex organic species were predicted to form abundantly on the grain surfaces according to the scenario of \citet{garrod06a} and \citet{garrod08a}.

The current paper extends the chemical analysis to a full gas-phase network, including freeze-out onto and evaporation from dust grains, as well as basic grain-surface hydrogenation reactions. Combining semi-analytical density and velocity structures with detailed temperature profiles from full radiative transfer calculations, our aim is to bridge the gap between 1D chemical models of collapsing cores and 1+1D or 2D chemical models of T Tauri and Herbig Ae/Be disks \citep{bergin07a}. One of the key questions is whether the chemical composition of such disks results mainly from chemical processing during the collapse or whether it is determined by in situ processing after the disk has formed.

As reviewed by \citet{difrancesco07a} and \citet{bergin07b}, the chemistry of pre-stellar cores is well understood. Because of the low temperatures and the moderately high densities, many molecules are depleted from the gas by freezing out onto the cold dust grains. The main ice constituent is \w, showing abundances of $\sim$\ten{-4} relative to gas-phase \mh{} \citep{tielens91a,pontoppidan04a}. Other abundant ices are \cdo{} and CO \citep{gibb04a,oberg11b}. Correspondingly, the observed gas-phase abundances of \w{}, CO and \cdo{} in pre-stellar cores are low \citep{snell00c,bacmann02a,caselli10a}. Nitrogen-bearing species like \mn{} and \amh{} are generally less depleted than carbon- and oxygen-bearing ones \citep{tafalla04a}, partly because they require a longer time to be formed in the gas and therefore have not yet had a chance to freeze out \citep{aikawa01a,difrancesco07a}, and partly because CO is no longer present as a destructive agent \citep{rawlings92a}. The observed depletion factors are well reproduced with 1D chemical models \citep{bergin97b,lee04a}.

The collapse phase is initially characterised by a gradual warm-up of the material, resulting in the evaporation of the ices according to their respective binding energies \citep{vandishoeck98a,jorgensen04c}. The higher temperatures also drive a rich chemistry, especially if it gets warm enough to evaporate \w{} and organic species like \meoh{} and \mf{} \citep{charnley92a}. Spherical models of the chemical evolution during the collapse phase \citep[e.g.,][]{rodgers03a,lee04a,garrod06a,aikawa08a,garrod08a} are successful at explaining the observed abundances at scales of several thousand AU, where the envelope is still close to spherically symmetric, but they cannot make the transition from the 1D spherically symmetric envelope to the 2D axisymmetric circumstellar disk. Recently, \citet{vanweeren09a} followed the chemical evolution within the framework of a 2D hydrodynamical simulation and obtained a reasonable match with observations of protostars. Even though their primary focus was still on the envelope and not on the disk, they did show how important it is to treat the chemical evolution during low-mass star formation in more than one dimension.

Once the phase of active accretion from the envelope comes to an end, the circumstellar disk settles into the comparatively static T Tauri or Herbig Ae/Be phase. So far only some simple molecules have been detected in disks \citep{dutrey97a,kastner97a,thi04a,oberg10a,oberg11a}, but the inventory is expected to grow now that the Atacama Large Millimeter Array (ALMA) is becoming operational. Chemically, disks can be divided into three layers: a cold zone near the midplane, a warm molecular layer at intermediate altitudes, and a photon-dominated region at the surface \citep{bergin07a}. However, it is unknown to what extent the chemical composition of the disk is affected by the chemical evolution in the collapsing core at earlier times, or how much chemical processing has been experienced by material ending up in planetary and cometary building blocks.

This paper aims to provide an important step towards filling in this missing link by following the chemical evolution all the way from a pre-stellar cloud core to a circumstellar disk in two spatial dimensions. The physical and chemical models are described in Sects.\ \ref{sec:collmod} and \ref{sec:chemnet}. The chemistry during the pre-collapse phase is briefly reviewed in Sect.\ \ref{sec:precoll}, followed by an extensive discussion of the collapse-phase chemistry in Sect.\ \ref{sec:coll}. The abundances resulting from the collapse phase are compared to in situ processing in a static disk in Sect.\ \ref{sec:local}. Sections \ref{sec:comets} and \ref{sec:cav} discuss the implications of our results for the origin of comets and address some caveats in the model. Conclusions are drawn in Sect.\ \ref{sec:conc}.


\section{Collapse model}
\label{sec:collmod}


\subsection{Step-wise summary}
\label{subsec:msteps}
\begin{figure}
\resizebox{\hsize}{!}{\includegraphics{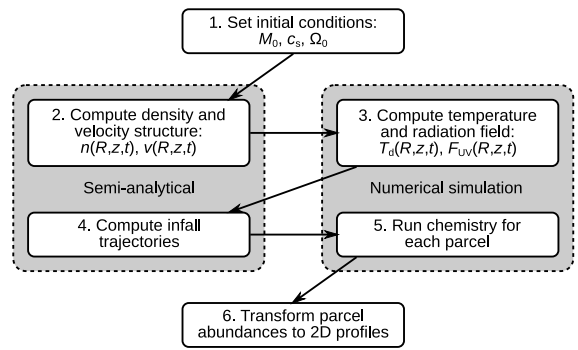}}
\caption{Step-wise summary of our 2D axisymmetric collapse model. Steps 2 and 4 are semi-analytical, while steps 3 and 5 consist of detailed numerical simulations.}
\label{fig:msteps}
\end{figure}

Our semi-analytical collapse model is described in detail in \vddd{} and summarised in \fig{msteps}. The model starts with a singular isothermal sphere characterised by a total mass $M_0$, an effective sound speed $\cs$, and a uniform rotation rate $\Omega_0$. As soon as the collapse starts, at $t=0$, the rotation causes the infalling material to be deflected towards the equatorial midplane. This breaks the spherical symmetry, so the entire model is run as a two-dimensional axisymmetric system. The 2D density and velocity profiles follow the solutions of \citet{shu77a}, \citet{cassen81a} and \citet{terebey84a} for an inside-out collapse with rotation. After the disk is first formed at the midplane, it evolves by ongoing accretion from the collapsing core and by viscous spreading to conserve angular momentum \citep{shakura73a,lyndenbell74a}.

Taking the 2D density profiles from step 2, and adopting the appropriate size and luminosity for the protostar \citep{adams86a,young05a}, the next step consists of computing the dust temperature at a number of time steps. This is done with the radiative transfer code RADMC \citep{dullemond04a}, which takes a 2D axisymmetric density profile but follows photons in all three dimensions. RADMC also computes the full radiation spectrum at each point in the axisymmetric disk and remnant envelope, as required for the photon-driven reactions in our chemical network (Sects.\ \ref{subsec:rad} and \ref{subsec:photo}). The gas temperature is set equal to the dust temperature throughout the disk and the envelope. This is a poor assumption in the surface of the disk and the inner parts of the envelope \citep{kamp04a,jonkheid04a}, the consequences of which are addressed in Sect.\ \ref{sec:cav}. As argued in \vddd, the disk-envelope accretion shock is not important for the chemical evolution of the material in the disk at the end of the collapse phase. It is therefore not incorporated in our model.

Given the dynamical nature of the collapse, it is easiest to solve the chemistry in a Lagrangian frame. In \vddd, the envelope was populated with several thousand parcels at $t=0$ and these were followed in towards the disk and star. The method is reversed here: a regular grid of parcels is defined at the end of the collapse and the parcels are followed backwards in time to their position at $t=0$. The reason for doing so is that the 2D abundance profiles at the end of the collapse, when the disk is fully formed, are of greater interest than those at the onset of collapse. In either case, step 4 of the model produces a set of infall trajectories with densities, temperatures and UV intensities as a function of time and position. These data are required for the next step: solving the time-dependent chemistry for each individual parcel. Although the parcels are followed backwards in time to get their trajectories, the chemistry is computed in the normal forward direction. The last step of our model consists of transforming the abundances from the individual parcels back into 2D axisymmetric profiles at any time step of interest.

In \vddd{}, the model was run for a grid of initial conditions. In the current paper, the analysis is limited to our standard set of parameters: $M_0=1.0$ $M_\odot$, $\cs=0.26$ km \ps{} and $\Omega_0=10^{-14}$ \ps. The viscosity parameter $\alpha$ is fixed at 0.01 \citep{hartmann98a}. Section \ref{sec:comets} contains a brief discussion on how the results may change for other parameter values.


\subsection{Differences with \vddd}
\label{subsec:diff}
The current version of the model contains several improvements over the version used in \vddd, described in detail by \citet{visser10a}. Most importantly, the model now correctly treats the problem of sub-Keplerian accretion onto a 2D disk. Material falling onto the disk along an elliptic orbit has sub-Keplerian angular momentum, so it exerts a torque on the disk that results in an inward push. Several solutions are available \citep[e.g.,][]{cassen81a,hueso05a}, but these are not suitable for our 2D model. The ad-hoc solution from \vddd{} provided the appropriate qualitative physical correction -- increasing the inward radial velocity of the disk material -- but it did not properly conserve angular momentum. The updated model uses a new, fully consistent solution, derived directly from the equations for the conservation of mass and angular momentum \citep{visser10a}. It results in disks that are typically a factor of a few smaller than those obtained previously, but the disk masses are mostly unchanged. The new disks are a few degrees colder in the inner part and warmer in the outer part, which may further affect the chemistry.

Other changes to the model include the definition of the disk-envelope boundary and the shape of the outflow cavity. In \vddd, the disk-envelope boundary was defined as the surface where the density of the infalling envelope material equals that of the disk. The model now uses the surface where the ram pressure of the infalling material equals the thermal pressure of the disk \citep{visser10a}, providing a more physically correct description of where material becomes part of the disk. The collapse solution of \citet{terebey84a} does not produce an outflow cavity, so one is added manually. The original model used a cavity with straight walls, but observations and theoretical predictions show that curved walls are more appropriate \citep{velusamy98a,canto08a}. The adopted description of the outflow wall is now
\begin{equation}
\label{eq:outflow}
z = (0.191\ {\rm AU})\left(\frac{t}{\tacc}\right)^{-3}\left(\frac{R}{{\rm AU}_{}}\right)^{1.5}\,,
\end{equation}
with $R$ and $z$ in cylindrical coordinates and $\tacc=M_0/\dot{M}$ the time required for the entire envelope to accrete onto the star and disk. The $t^{-3}$ dependence is chosen so that the outflow starts narrow and becomes wider as the collapse proceeds. The full opening angle at $\tacc$ is 33.6$^\circ$ at $z=1000$ AU and 15.9$^\circ$ at $z=10\,000$ AU. There are some streamlines in the infalling envelope that hit the cavity wall, but the mass flowing along these lines is insignificant ($\sim$1\%) compared to the mass accreting onto the star and the disk.


\subsection{Radiation field}
\label{subsec:rad}
Photodissociation and photoionisation by ultraviolet (UV) radiation are important processes in the hot inner core and in the surface layers of the disk. The temperature and luminosity of the protostar change in time (\vddd), so neither the strength nor the spectral shape of the radiation field is constant. In addition, the spectral shape changes as the radiation passes through the disk and remnant envelope.

The most accurate way of obtaining the time- and location-dependent photorates is to multiply the cross section for each reaction by the UV field at each grid point. The latter can be computed from 2D radiative transfer at high spectral resolution. As this is too computationally demanding, several approximations have to be made. The first one is to assume that the wavelength-dependent attenuation of the radiation field by the dust in the disk and envelope can be represented by a single factor $\gamma$ for each reaction \citep{vandishoeck06a}. The rate coefficient for a given photoreaction at spatial coordinates $r$ and $\theta$ and at time $t$ can then be expressed as
\begin{equation}
\label{eq:phrate}
k_{\rm ph}^{}(r,\theta,t) = k_{\rm ph}^\ast\left(\frac{r}{R_\ast}\right)^{-2}\exp(-\gamma\av)\,,
\end{equation}
with $\av$ the visual extinction (see below). The unshielded rate coefficient is calculated at the stellar surface ($k_{\rm ph}^\ast$) by multiplying the cross section of the reaction by the blackbody flux at the effective temperature of the protostar, which varies between 4500 and 5500 K for the duration of the collapse (\vddd). Any excess UV from the disk-star boundary is not included. The term $(r/R_\ast)^{-2}$ accounts for the geometrical dilution of the radiation from the star across a distance $r$. The factor $\gamma$ is discussed in Sect.\ \ref{subsec:photo}.

In order to apply \eq{phrate}, the extinction towards each point ($r,\theta$) is calculated with RADMC at a low spectral resolution of one frequency point per eV\@. The local UV flux from RADMC at ($r,\theta$), $F_{\rm RADMC}$ (integrated from 6 to 13.6 eV), can be considered the product of the unattenuated flux and an extinction term,
\begin{equation}
\label{eq:taueff}
F_{\rm RADMC}(r,\theta,t) = F_0\exp(-\tau_{\rm UV,eff}) = F_\ast\left(\frac{r}{R_\ast}\right)^{-2}\exp(-\tau_{\rm UV,eff})\,.
\end{equation}
$F_\ast$ is the time-dependent UV flux at the stellar surface. The effective UV optical depth, $\tau_{\rm UV,eff}$, is an average over the many possible paths a photon can follow from the source to the point ($r,\theta$). The large number of photons propagated through the grid by RADMC (typically \ten{5}) ensures a statistical sampling of all possible trajectories. The UV optical depth is converted to the visual extinction $\av$ through the standard relationship $\av=\tau_{\rm UV,eff}/3.02$ \citep{bohlin78a}.


\section{Chemical network}
\label{sec:chemnet}
The basis of our chemical network is the UMIST06 database \citep{woodall07a} as modified by \citet{bruderer09a}, except that X-ray chemistry is not included. The modified network includes charge-exchange reactions between positive ions and grains \citep{maloney96a}, which can be important in the disk's inner few AU\@. The cosmic-ray ionisation rate of \mh{} is set to \scit{5}{-17} \ps{} \citep{dalgarno06a}. Attenuation of cosmic rays at column densities of more than $\sim$100 g \pcs{} \citep{umebayashi81a} is not taken into account, even though this can become relevant for the midplane chemistry in the inner few AU of massive disks. The network contains 162 neutral species, 251 cations and six anions, built up out of 18 elements. The initial composition is fully atomic, except that hydrogen starts as \mh. Elemental abundances are adopted from \citet{aikawa08a}. Additional values are taken from \citet{bruderer09a}, but reduced by a factor of 100 to have a consistent set of low-metal abundances. \tb{elabun} lists the elemental abundances relative to the total hydrogen nucleus density: $\nh=n({\rm H})+2n(\mhm)$. 

\begin{table}
\caption{Elemental abundances: $x(X)=n(X)/\nh$.}
\label{tb:elabun}
\centering
\begin{tabular}{ccccc}
\hline\hline
Species & Abundance & & Species & Abundance \\
\hline
\mh & 5.00(-1) & & Si & 2.74(-9) \\
He  & 9.75(-2) & & P  & \pz2.16(-10) \\
C   & 7.86(-5) & & S  & 9.14(-8) \\
N   & 2.47(-5) & & Cl & 1.00(-9) \\
O   & 1.80(-4) & & Ar & 3.80(-8) \\
Ne  & 1.40(-6) & & Ca & 2.20(-8) \\
Na  & 2.25(-9) & & Cr & 4.90(-9) \\
Mg  & 1.09(-8) & & Fe & 2.74(-9) \\
Al  & 3.10(-8) & & Ni & 1.80(-8) \\
\hline
\end{tabular}
\tablefoot{The notation $a(b)$ means \scit{a}{b}.}
\end{table}

In order to set the chemical composition at the onset of collapse ($t=0$), the initially atomic gas is evolved for a period of 1 Myr at $\nh=\scim{8}{4}$ \pcc{} and $T_{\rm g}=\td=10$ K\@. The visual extinction during this pre-collapse phase is set to 100 mag to disable all photoprocesses, except for a minor contribution from cosmic-ray--induced photons. The resulting solid and gas-phase abundances are consistent with those observed in pre-stellar cores \citep[e.g.,][]{difrancesco07a}. They form the initial conditions for the collapse phase for all infalling parcels. In the remainder of this paper, $t=0$ always refers to the onset of collapse, following the 1 Myr pre-collapse phase here described.


\subsection{Photodissociation and photoionisation}
\label{subsec:photo}
Photodissociation and photoionisation by UV radiation are important processes in the inner disk and inner envelope. Their rates are given by \eq{phrate}, using the effective extinction from \eq{taueff}. The unshielded rates at the stellar surface ($k_{\rm ph}^\ast$) are calculated for a blackbody spectrum at the time-dependent stellar temperatures from \vddd. The required cross sections are taken from our freely available database.\footnote{http://www.strw.leidenuniv.nl/$\sim$ewine/photo} Values for the extinction factor $\gamma$ from \eq{phrate} were tabulated by \citet{vandishoeck06a} for three different types of radiation fields, of which the 4000 K blackbody is most appropriate for our model (Sect.\ \ref{subsec:rad}). The final output of this procedure is a 3D array (two spatial coordinates and time) with rate coefficients ($k_{\rm ph}$) for each photoreaction. When computing the infall trajectories for individual parcels (step 4 in \fig{msteps}), the rate coefficients at all points along each trajectory are obtained from a linear interpolation.

The photodissociation of \mh{} and CO requires some special treatment. Both processes occur exclusively through discrete absorption lines, so self-shielding plays an important role. The amount of shielding for \mh{} is a function of the \mh{} column density; for CO, it is a function of both the CO and the \mh{} column density, because some CO lines are shielded by \mh{} lines. The effective UV extinction from \eq{taueff} can be converted into a total hydrogen column density through $\av=\tau_{\rm UV,eff}/3.02$ and $N_{\rm H}=\scim{1.59}{21}\av$ \pcs{} \citep{bohlin78a,diplas94a}. Most hydrogen along each photon path is in molecular form, so the effective $N(\mhm)$ towards each spatial grid point is $\sim$$0.5N_{\rm H}$. Equation (37) from \citet{draine96a} then gives the amount of self-shielding for \mh. The unshielded dissociation rate is computed according to the one-line approximation from \citet{vandishoeck87a}, scaled so that the rate is \scit{4.5}{-11} \ps{} in the standard \citet{draine78a} field. The photodissociation rate of CO is computed from our new shielding functions and cross sections \citep{visser09b}. Effective CO column densities for the self-shielding calculation are derived from the \mh{} column densities by assuming an average $N({\rm CO})/N(\mhm)$ ratio of \ten{-5}, accounting for the partial dissociation of CO along each photon path. The results are not sensitive to the exact choice of average CO abundance \citep{vanzadelhoff03a}.


\subsection{Gas-grain interactions}
\label{subsec:grains}
All neutral species other than H, \mh{}, He, Ne and Ar are allowed to freeze out onto the dust according to \citet{charnley01a}. In cold, dense environments -- such as our model cores before the onset of collapse -- observations show \w, CO and \cdo{} to be the most abundant ices \citep{boogert11a}. As the temperature rises during the collapse, the ices evaporate according to their respective binding energies. However, the presence of non-volatile species like \w{} prevents the more volatile species like CO and \cdo{} from evaporating entirely: some CO and \cdo{} gets trapped in the \w{} ice \citep{hasegawa93a,fayolle11a}. The results from \vddd{} suggested that this is required to explain the presence of CO in solar-system comets.

\begin{table}
\caption{Pre-exponential factors and binding energies for selected species in our network.}
\label{tb:ebind}
\vspace{-0.1cm}
\centering
\begin{tabular}{cccc}
\hline\hline
\rule{0pt}{1em}Species & $\nu$ (\ps) & $E_{\rm b}/k$ (K) & Reference \\
\hline
\multicolumn{1}{l}{\rule{0pt}{1em}\,\,\ \w}  & \scit{1}{15} & 5773           & \multicolumn{1}{l}{\citet{fraser01a}} \\
\multicolumn{1}{l}{\,\,\ CO}  & \scit{9}{11} & 855 & \multicolumn{1}{l}{\citet{bisschop06a}} \\
\multicolumn{1}{l}{\,\,\ \mn} & \scit{1}{11} & 800 & \multicolumn{1}{l}{\citet{bisschop06a}} \\
\multicolumn{1}{l}{\,\,\ \mo} & \scit{9}{11} & 912 & \multicolumn{1}{l}{\citet{acharyya07a}} \\
\hline
\end{tabular}
\end{table}

Incorporating ice trapping in a chemical network is a non-trivial task \citep{viti04a,fayolle11a}, so the process is ignored for now. Instead, desorption of all species is treated according to the zeroth-order rate equation
\begin{equation}
\label{eq:thdes}
R_{\rm thdes}(X) = 4\pi a_{\rm gr}^2n_{\rm gr}^{}f(X)\nu(X)N_{\rm ss}\exp\left[-\frac{E_{\rm b}(X)}{k\td}\right]\,,
\end{equation}
where $\td$ is the dust temperature, $a_{\rm gr}=0.1$ \micron{} the typical grain radius, and $n_{\rm gr}=\scim{1}{-12}\nh$ the grain number density. The canonical pre-exponential factor, $\nu$, for \emph{first}-order desorption is \scit{2}{12} \ps{} \citep{sandford93a}, which is used for all ices except the four listed in \tb{ebind}. In order to get the \emph{zeroth}-order pre-exponential factor, $\nu$ is multiplied by $N_{\rm ss}$, the number of binding sites per unit grain surface. Assuming $N_{\rm ss}=\scim{8}{14}$ \pcs{}, the number of binding sites per grain is $N_{\rm b}=\scim{1}{6}$ for our 0.1 \micron{} grains. The binding energies of species other than the four in \tb{ebind} are set to the values tabulated by \citet{sandford93a} and \citet{aikawa97a}. Species for which the binding energy is unknown are assigned the binding energy and the pre-exponential factor of \w\@. The dimensionless factor $f$ in \eq{thdes} ensures that each species desorbs according to its solid-phase abundance, and changes the overall desorption behaviour from zeroth to first order when there is less than one monolayer of ice:
\begin{equation}
\label{eq:desf}
f(X) = \frac{n_{\rm s}(X)}{\max(n_{\rm ice},N_{\rm b}n_{\rm gr})}\,,
\end{equation}
with $n_{\rm s}(X)$ the number density of ice species $X$, $N_{\rm b}=10^6$ the typical number of binding sites per grain, and $n_{\rm ice}$ the total number density (per unit volume of cloud or disk) of all ice species combined. Section \ref{sec:cav} briefly discusses how our results would change if trapping were included.

In addition to thermal desorption, our model includes desorption induced by UV photons. Laboratory experiments on the photodesorption of \w, CO and \cdo{} all produce a yield of $Y\approx10^{-3}$ molecules per grain per incident UV photon \citep{oberg07a,oberg09a}, while the yield for \mn{} is an order of magnitude lower \citep{oberg09b}. For all other ice species in our network, whose photodesorption yields have not yet been determined experimentally or theoretically, the yield is set to \ten{-3}. The photodesorption rate is
\begin{equation}
\label{eq:phdes}
R_{\rm phdes}(X) = \pi a_{\rm gr}^2n_{\rm gr}^{}f(X)Y(X)F_0\exp(-\tau_{\rm UV,eff})\,,
\end{equation}
with $f$ the same factor as for thermal desorption. The unattenuated UV flux ($F_0$) and the effective UV extinction ($\tau_{\rm UV,eff}$) follow from \eq{taueff}. Photodesorption occurs even in strongly shielded regions because of cosmic-ray--induced photons, which is accounted for by setting a lower limit of \ten{4} \pcs{} \ps{} to the product $F_0\exp(-\tau_{\rm UV,eff})$ \citep{shen04a}.

The chemical reactions in our model are not limited to the gas phase. As usual, the network includes the grain-surface formation of \mh{} \citep{black87a}. Inspired by \citet{bergin97b} and \citet{hollenbach09a}, it also includes the grain-surface hydrogenation of C to \meh, N to \amh, O to \w, and S to \hhs\@. The hydrogenation is done one H atom at a time and is always in competition with thermal and photon-induced desorption. The formation of \meh, \amh, \w{} and \hhs{} does not have to start with the respective atom freezing out. For instance, CH freezing out from the gas is also subject to hydrogenation on the grain surface. The rate of each hydrogenation step is taken to be the adsorption rate of H from the gas multiplied by the probability that the H atom encounters the atom or molecule $X$ to hydrogenate:
\begin{equation}
\label{eq:hydro}
R_{\rm hydro}(X) = \pi a_{\rm gr}^2n_{\rm gr}^{}n({\rm H})f'(X)\sqrt{\frac{8kT_{\rm g}}{\pi m_{\rm p}}}\,
\end{equation}
with $T_{\rm g}$ the gas temperature. The factor $f'$ serves a similar purpose as the factor $f$ in Eqs.\ (\ref{eq:thdes}) and (\ref{eq:phdes}). Since the hydrogenation is assumed to be near-instantaneous as soon as the H atom meets $X$ before $X$ desorbs, $X$ is assumed to reside always near the top layer of the ice. Hence, the abundance of solid $X$ should not be compared to the total amount of ice (as in $f$), but to the combined abundance of other ice species that can be hydrogenated:
\begin{equation}
\label{eq:fhydro}
f'(X) = \frac{n_{\rm s}(X)}{\max(n_{\rm hydro},N_{\rm b}n_{\rm gr})}\,,
\end{equation}
with $n_{\rm hydro}$ the sum of the solid abundances of the eleven species subject to hydrogenation: C, CH, \chh, \chhh, N, NH, \nhh, O, OH, S and SH\@. The main effect of this hydrogenation scheme is to build up an ice mixture of simple saturated molecules during the pre-collapse phase, as is found observationally \citep[e.g.,][]{tielens91a,gibb04a}.

In reality, grain-surface chemistry is not limited to the simple hydrogenation steps included here. For example, CO can react with H to form \fmh{} and \meoh{} or with OH to form \cdo{} \citep{fuchs09a,ioppolo11a}, and \mo{} can be hydrogenated to \w{} \citep{ioppolo08a}. Grains also play an important role in the formation of more complex species \citep{garrod06a,garrod08a}. However, none of these reactions can be implemented as easily as the hydrogenation of C, N, O and S\@. In addition, the main focus of this paper is on simple molecules whose abundances can be well explained with conventional gas-phase chemistry. The effects of more complex grain-surface chemistry will be explored in detail in a follow-up paper.


\section{Results from the pre-collapse phase}
\label{sec:precoll}
This section, together with the next two, contains the results from the gas-phase chemistry in our collapse model. The chemistry during the pre-collapse phase is briefly discussed first. The chemistry during the collapse is analysed in detail for one particular parcel in Sect.\ \ref{subsec:1parcel} and then generalised to others in Sect.\ \ref{subsec:other}. Finally, the collapse chemistry is compared to a static disk model in Sect.\ \ref{sec:local}. The results in the current section are all consistent with available observational constraints on pre-stellar cores \citep[e.g.,][]{difrancesco07a}.

During the 1.0 Myr pre-collapse phase, most of the initial atomic oxygen freezes out and is hydrogenated to \w{} ice. Meanwhile, \mh{} is ionised by cosmic rays. The resulting \hhp{} reacts with \mh{} to give \hhhp, which reacts with the remaining atomic O to ultimately form \mo:
\begin{align}
{\rm O}\ +\ {\rm H}_3^+\ \to\ &\ {\rm OH}^+\ +\ {\rm H}_2\,, \label{eq:o+h3p} \\
{\rm OH}^+\ +\ {\rm H}_2\ \to\ &\ {\rm H}_2{\rm O}^+\ +\ {\rm H}\,, \label{eq:ohp+h2} \\
{\rm H}_2{\rm O}^+\ +\ {\rm H}_2\ \to\ &\ {\rm H}_3{\rm O}^+\ +\ {\rm H}\,, \label{eq:h2op+h2} \\
{\rm H}_3{\rm O}^+\ +\ {\rm e}^-\ \to\ &\ {\rm OH}\ +\ {\rm H}_2/2{\rm H}\,, \label{eq:h3op+e} \\
{\rm OH}\ +\ {\rm O}\ \to\ &\ {\rm O}_2\ +\ {\rm H}\,. \label{eq:oh+o}
\end{align}
The \mo{} thus produced freezes out for the most part. At the onset of collapse, the four major oxygen reservoirs are \w{} ice (44\%), CO ice (34\%), \mo{} ice (16\%) and NO ice (3\%).

The oxygen chemistry is tied closely to the carbon chemistry through CO\@. It is initially formed in the gas phase from \chh, which in turn is formed from atomic C:
\begin{align}
{\rm C}\ +\ {\rm H}_2\ \to\ &\ {\rm CH}_2\ +\ h\nu\,, \label{eq:c+h2} \\
{\rm CH}_2\ +\ {\rm O}\ \to\ &\ {\rm CO}\ +\ 2{\rm H}\,. \label{eq:ch2+o}
\end{align}
Another early pathway from C to CO is powered by \hhhp{} and goes through an \hcop{} intermediate:
\begin{align}
{\rm C}\ +\ {\rm H}_3^+\ \to\ &\ {\rm CH}^+\ +\ {\rm H}_2^{}\,, \label{eq:c+h3p} \\
{\rm CH}^+\ +\ {\rm H}_2^{}\ \to\ &\ {\rm CH}_2^+\ +\ {\rm H}\,, \label{eq:chp+h2} \\
{\rm CH}_2^+\ +\ {\rm H}_2^{}\ \to\ &\ {\rm CH}_3^+\ +\ {\rm H}\,, \label{eq:ch2p+h2} \\
{\rm CH}_3^+\ +\ {\rm O}\ \to\ &\ {\rm HCO}^+\ +\ {\rm H}_2^{}\,, \label{eq:ch3p+o} \\
{\rm HCO}^+\ +\ {\rm C}\ \to\ &\ {\rm CO}\ +\ {\rm CH}^+\,. \label{eq:hcop+c}
\end{align}
The formation of CO through these two pathways accounts for most of the pre-collapse processing of carbon: at $t=0$, 82\% of all carbon has been converted into CO, of which 97\% has frozen out onto the grains. Most of the remaining carbon is present as \meh{} ice (14\% of all C), formed from the rapid grain-surface hydrogenation of atomic C.

The initial nitrogen chemistry consists mostly of converting atomic N into \amh{}, \mn{} and NO\@. The first of these is formed on the grains after freeze-out of N, in the same way that \w{} and \meh{} are formed from adsorbed O and C\@. The model results show two pathways leading to \mn. The first one starts with the cosmic-ray dissociation of \mh{} into H$^+$ and H$^-$, which \citet{cravens78a} estimated to have a 0.015\% chance of happening for each cosmic-ray ionisation event:
\begin{align}
{\rm H}_2\ +\ \zeta\ \to\ &\ {\rm H}^+\ +\ {\rm H}^-\,, \label{eq:h2+CR} \\
{\rm N}\ +\ {\rm H}^-\ \to\ &\ {\rm NH}\ +\ {\rm e}^-\,, \label{eq:n+hm} \\
{\rm NH}\ +\ {\rm N}\ \to\ &\ {\rm N}_2\ +\ {\rm H}\,. \label{eq:nh+n}
\end{align}
The other pathway couples the nitrogen chemistry to the carbon chemistry. It starts with Reactions (\ref{eq:c+h3p})--(\ref{eq:ch2p+h2}) to form \mep, followed by
\begin{align}
{\rm CH}_3^+\ +\ {\rm e}^-\ \to\ &\ {\rm CH}\ +\ {\rm H}_2^{}/2{\rm H}\,, \label{eq:ch3p+e} \\
{\rm CH}\ +\ {\rm N}\ \to\ &\ {\rm CN}\ +\ {\rm H}\,, \label{eq:ch+n} \\
{\rm CN}\ +\ {\rm N}\ \to\ &\ {\rm N}_2\ +\ {\rm C}\,. \label{eq:cn+n}
\end{align}
The nitrogen chemistry is also tied to the oxygen chemistry, forming NO out of N and OH:
\begin{equation}
\label{eq:n+oh}
{\rm N}\ +\ {\rm OH}\ \to\ {\rm NO}\ +\ {\rm H}\,,
\end{equation}
with OH formed by \rx{h3op+e}. Nearly all of the \mn{} and NO formed during the pre-collapse phase freezes out. At $t=0$, solid \mn, solid \amh{} and solid NO account for 41, 32 and 22\% of all nitrogen.


\section{Results from the collapse phase}
\label{sec:coll}
The collapse-phase chemistry is run for the standard set of model parameters from \vddd{}: $M_0=1.0$ $M_\odot$, $\cs=0.26$ km \ps{} and $\Omega_0=10^{-14}$ \ps. The core collapses in $\tacc=\scim{2.52}{5}$ yr to form a 0.76 $M_\odot$ star surrounded by a compact disk (47 AU outer radius) of 0.13 $M_\odot$. At $t=\tacc$, the radius at which \w{} freezes out (the snowline) lies at 5.1 AU. The \citet{toomre64a} $Q$ parameter ranges from 130 at 0.1 AU to a minimum of 1.4 at 28 AU, so the disk remains gravitationally stable by a small margin.


\subsection{One single parcel}
\label{subsec:1parcel}
\begin{figure*}
\resizebox{\hsize}{!}{\includegraphics{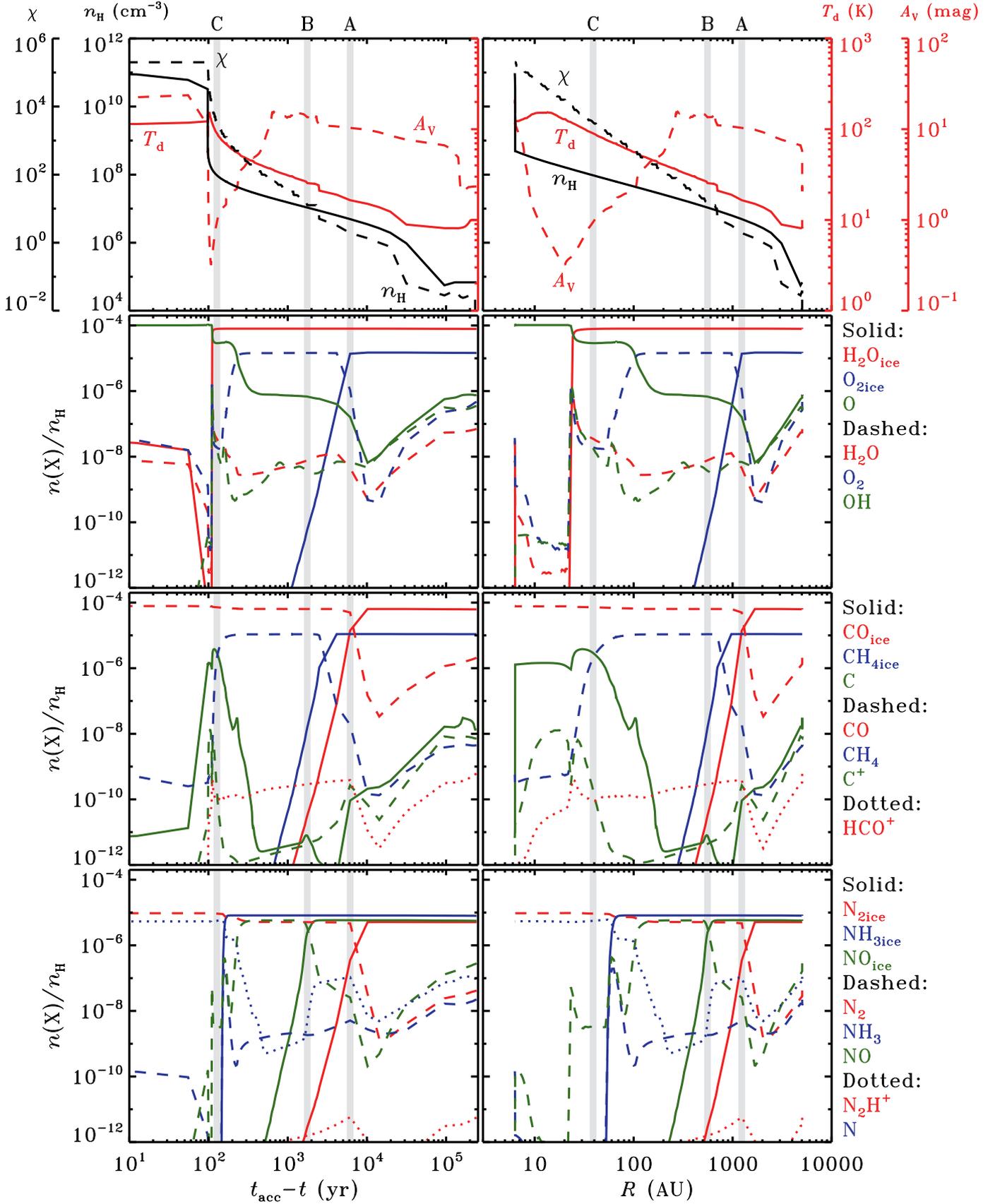}}
\caption{Physical conditions ($\nh$, $\td$, $\chi$ and $\av$) and abundances of the main oxygen-, carbon- and nitrogen-bearing species for the single parcel from Sect.\ \ref{subsec:1parcel}, as function of time before the end of the collapse (left) and as function of horizontal position during the collapse (right). The grey bars correspond to the points A, B and C from \fig{1over}. In each panel, the parcel essentially moves from right to left.}
\label{fig:1abun}
\end{figure*}

The chemical evolution is first discussed in detail for one particular infalling parcel of material. It starts near the edge of the cloud core, at $r=6710$ AU from the center and $\theta=48.8^\circ$ degrees from the $z$ axis ($R=5050$ AU, $z=4420$ AU). Its trajectory terminates at $t=\tacc$ in the inner part of the disk, at $R=6.3$ AU and $z=2.4$ AU, about 0.2 AU below the surface. The physical conditions encountered along the trajectory are plotted in \fig{1abun}. The radiation field is characterised by the unattenuated flux and the visual extinction, where the former is expressed as a scaling factor relative to the average flux in the interstellar medium (ISM): $\chi=F_0/F_{\rm ISM}$, with $F_{\rm ISM}=\scim{8}{7}$ photons \pcs{} \ps{} \citep{draine78a}. This figure also shows the abundances of the main oxygen-, carbon- and nitrogen-bearing species. The four panels in the right column are regular plots as function of $R$ (the coordinate along the midplane), tracing the parcel's trajectory from $t=0$ ($R=5050$ AU) to $t=\tacc$ ($R=6.3$ AU). The infall velocity of the parcel increases as it gets closer to the star, so the physical conditions and chemical abundances change more rapidly at later times. Hence, the panels in the left column are plotted as a function of $\tacc-t$: the time remaining before the end of the collapse phase. In each individual panel, the parcel essentially moves from right to left.

\begin{figure*}[t!]
\resizebox{\hsize}{!}{\includegraphics{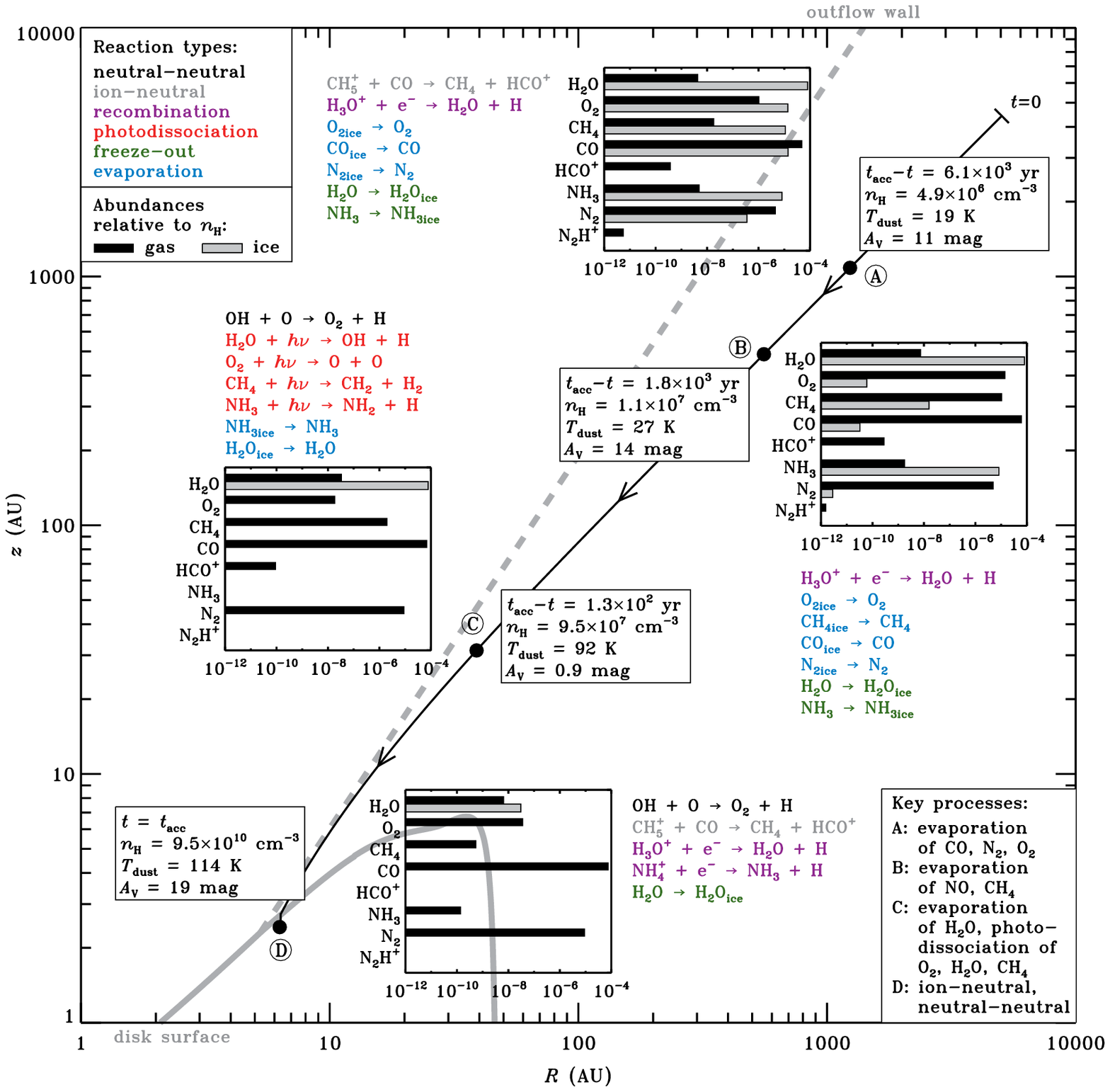}}
\caption{Overview of the chemistry along the infall trajectory (solid black curve) of the single parcel from Sect.\ \ref{subsec:1parcel}. The solid and dashed grey lines denote the surface of the disk and the outflow wall, both at $t=\tacc=\scim{2.52}{5}$ yr. Physical conditions, abundances (black bars: gas; grey bars: ice) and key reactions are indicated at four points (A, B, C and D) along the trajectory. The key processes governing the overall chemistry at each point are listed in the bottom right. The type of each reaction is indicated by colour, as listed in the top left.}
\label{fig:1over}
\end{figure*}

\figg{1over} presents a schematic overview of the parcel's chemical evolution. It shows the infall trajectory of the parcel and the abundances of several species at four points along the trajectory. The physical conditions and the key reactions controlling those abundances are also listed. Most abundance changes for individual species are related to one specific chemical event, such as the evaporation of CO or the photodissociation of \w\@. The remainder of this subsection discusses the abundance profiles from \fig{1abun} and explains them in the context of \fig{1over}.

\subsubsection{Oxygen chemistry}
\label{sss:1oxygen}
At the onset of collapse ($t=0$), the main oxygen reservoir is solid \w{} at an abundance of \scit{8}{-5} relative to $\nh$ (Sect.\ \ref{sec:precoll}). The abundance remains constant until the parcel gets to point C in \fig{1over}, where the temperature is high enough for the \w{} ice to evaporate. The parcel is now located close to the outflow wall, so the stellar UV field is only weakly attenuated ($\av=0.9$ mag). Hence, the evaporating \w{} is immediately photodissociated into H and OH, which in turn is dissociated into O and a second H atom. At $R=17$ AU (23 AU inside of point C), the dust temperature is 150 K and all solid \w{} is gone. Moving in further, the parcel enters the surface layers of the disk and is quickly shielded from the stellar radiation ($\av=10$--20 mag). The temperature decreases at the same time to 114 K, allowing some \w{} ice to reform. The final \w{} ice abundance at $\tacc$ (point D) is \scit{4}{-8}.

The dissociative recombination of \ohhhp{} (formed by Reactions (\ref{eq:o+h3p})--(\ref{eq:h2op+h2})) initially maintains \w{} in the gas at an abundance of \scit{7}{-8} at $t=0$. Following the sharp increase in the overall gas density at $t=\scim{2.1}{5}$ yr (\fig{1abun}), the freeze-out rate increases and the gas-phase \w{} abundance goes down to \scit{3}{-10} at point A in \fig{1over}. Moving on towards point B, the evaporation of \mo{} from the grains enables a new \w{} formation route:
\begin{align}
{\rm O}_2\ +\ {\rm C}^+\ \to\ &\ {\rm CO}\ +\ {\rm O}^+\,, \label{eq:o2+cp} \\
{\rm O}^+\ +\ {\rm H}_2\ \to\ &\ {\rm OH}^+\ +\ {\rm H}\,, \label{eq:op+h2}
\end{align}
followed by Reactions (\ref{eq:ohp+h2}) and (\ref{eq:h2op+h2}) to give \ohhhp, which recombines with an electron to give \w\@. The \w{} abundance thus increases to \scit{3}{-9} at $R=300$ AU\@. Further in, at point C, solid \w{} comes off the grains as described above. However, photodissociation keeps the gas-phase abundance from growing higher than $\sim$\ten{-7}. Once all \w{} ice is gone at $R=17$ AU, the gas-phase abundance can no longer be sustained at \ten{-7} and it drops to \scit{3}{-12}. Some \w{} is eventually reformed as the parcel gets into the disk and is shielded from the stellar radiation, producing a final abundance of $\scim{9}{-9}$ relative to $\nh$.

Another main oxygen reservoir at $t=0$ is \mo{}. Its solid and gas-phase abundances in the model are \scit{1}{-5} and \scit{4}{-7}, consistent with available observational constraints \citep{goldsmith85a,goldsmith11a,fuente93a}. Gaseous \mo{} gradually continues to freeze out until it reaches a minimum gas-phase abundance of \scit{3}{-10} just inside of point A\@. The temperature at that time is 18 K, enough for \mo{} to slowly start evaporating thermally. The gas-phase abundance is up by a factor of ten by the time the dust temperature reaches 23 K, about halfway between points A and B\@. The evaporation is 99\% complete as the parcel reaches $R=460$ AU, about 140 AU inside of point B\@. The gas-phase abundance remains stable at \scit{1}{-5} for the next few hundred years. Then, as the parcel gets closer to the outflow wall and into a region of lower extinction, the photodissociation of \mo{} sets in and its abundance decreases to \scit{2}{-8} at point C\@. The evaporation and photodissociation of \w{} at that point enhances the abundances of OH and O, which react with each other to replenish some \mo. As soon as all the \w{} ice is gone, this \mo{} production channel quickly disappears and the \mo{} abundance drops to \scit{1}{-11}. Finally, when the parcel enters the top of the disk, \mo{} is no longer photodissociated and its abundance goes back up to \scit{4}{-8} at point D.

The abundance of gas-phase OH starts at \scit{3}{-7}. Its main formation pathway is initially the dissociative recombination of \ohhhp{} (\rx{h3op+e}), and its main destructors are O, N and \hhhp. The increase in total density at \scit{2.1}{5} yr speeds up the destruction reactions, and the OH abundance drops to \scit{3}{-10} at point B\@. The evaporation of solid OH then briefly increases the gas-phase abundance to \scit{1}{-8}. When all of the OH has evaporated at $R=300$ AU, the gas-phase abundance goes down again to \scit{5}{-10} over the next 150 AU\@. As the parcel continues towards and past point C, the OH abundance is boosted to a maximum of \scit{1}{-6} by the photodissociation of \w\@. The high abundance lasts only briefly, however. As the last of the \w{} evaporates and gets photodissociated, OH can no longer be formed as efficiently, and it is dissociated itself. At the end of the collapse, the OH abundance is $\sim$\ten{-14}.

The fifth main oxygen-bearing species is atomic O itself. Its abundance is \scit{7}{-7} at $t=0$ and \scit{1}{-4} at $t=\tacc$, accounting for respectively 0.4 and 56\% of the total amount of non-refractory oxygen. Starting from $t=0$, the O abundance remains constant during the first \scit{2.0}{5} yr of the collapse phase. The increasing overall density then speeds up the reactions with OH (forming \mo) and \hhhp{} (forming \ohp), as well as the adsorption onto the grains, and the O abundance decreases to a low of \scit{2}{-8} just before point A\@. The abundance goes back up due to the evaporation of CO and, at point B, of \mo{} and NO, with O formed from the following reactions:
\begin{align}
{\rm CO}\ +\ {\rm He}^+\ \to\ &\ {\rm C}^+\ +\ {\rm O}\ +\ {\rm He}\,, \label{eq:co+hep} \\
{\rm NO}\ +\ {\rm N}\ \to\ &\ {\rm N}_2\ +\ {\rm O}\,, \label{eq:no+n} \\
{\rm O}_2\ +\ {\rm CN}\ \to\ &\ {\rm OCN}\ +\ {\rm O}\,. \label{eq:o2+cn}
\end{align}
Heading on towards point C, the photodissociation of \mo, NO and \w{} further drives up the amount of atomic O to the aforementioned final abundance of \scit{1}{-4}.


\subsubsection{Carbon chemistry}
\label{sss:1carbon}
With solid and gas-phase abundances of \scit{6}{-5} and \scit{2}{-6} relative to $\nh$, CO is the main form of non-refractory carbon at the onset of collapse. CO is a very stable molecule and its chemistry is straightforward. The freeze-out process started during the pre-collapse phase continues up to $t=\scim{2.4}{5}$ yr, a few thousand years prior to reaching point A in \fig{1over}, where the dust temperature of 18 K results in CO evaporating again. As the parcel continues its inward journey and is heated up further, all solid CO rapidly disappears and the gas-phase abundance goes up to \scit{6}{-5} at point B\@. During the remaining part of the infall trajectory, the other main carbon-bearing species are all largely converted into CO\@. At the end of the collapse (point D), 99.8\% of all carbon and 44\% of all oxygen are locked up in CO\@.

The second most abundant carbon-bearing ice at the onset of collapse is \meh, at \scit{1}{-5} with respect to $\nh$. The gas-phase abundance of \meh{} begins at \scit{4}{-9}, about a factor of 2500 lower. At point A, the evaporation of CO provides the first increase in $x({\rm CH}_4)$ through a chain of reactions starting with the formation of \cp{} from CO and He$^+$. The successive hydrogenation of \cp{} produces \mehhp, which reacts with another CO molecule to form \meh:
\begin{align}
{\rm CO}\ +\ {\rm He}^+\ \to\ &\ {\rm C}^+\ +\ {\rm O}\ +\ {\rm He}\,, \label{eq:co+hep-2} \\
{\rm C}^+\ +\ {\rm H}_2^{}\ \to\ &\ {\rm CH}_2^+\ +\ h\nu\,, \label{eq:cp+h2} \\
{\rm CH}_2^+\ +\ {\rm H}_2^{}\ \to\ &\ {\rm CH}_3^+\ +\ {\rm H}\,, \label{eq:ch2p+h2-2} \\
{\rm CH}_3^+\ +\ {\rm H}_2^{}\ \to\ &\ {\rm CH}_5^+\,, \label{eq:ch3p+h2} \\
{\rm CH}_5^+\ +\ {\rm CO}\ \to\ &\ {\rm CH}_4^{}\ +\ {\rm HCO}^+\,. \label{eq:ch5p+co}
\end{align}
The \meh{} ice evaporates at point B, bringing the gas-phase abundance up to \scit{1}{-5}. So far, the abundances of \meh{} and CO are well coupled. The link is broken when the parcel reaches point C, where \meh{} is photodissociated, but CO is not. This difference arises from the fact that CO can only be dissociated by photons shortwards of 1076 \AA, while \meh{} can be dissociated out to 1450 \AA{} \citep{visser09b}. The 5300 K blackbody spectrum emitted by the protostar at this time is not powerful enough at short wavelengths to cause significant photodissociation of CO\@. \meh{}, on the other hand, is quickly destroyed. Its final abundance at point D is \scit{6}{-10}.

Neutral and ionised carbon show the same trends in their abundance profiles, with the former always more abundant by a few per cent to a few orders of magnitude. Both start the collapse phase at $\sim$\ten{-8} relative to $\nh$. The increase in total density at \scit{2.1}{5} yr speeds up the destruction reactions (mainly by OH and \mo{} for C and by OH and \mh{} for \cp), so the abundances go down to $x({\rm C})=\scim{4}{-10}$ and $x({\rm C}^+)=\scim{5}{-11}$ just outside point A\@. This is where CO begins to evaporate, and as a result, the C and \cp{} abundances increase again. As the parcel continues to fall in towards point B, the evaporation of \mo{} and NO and the increasing total density cause a second drop in C and \cp. Once again, though, the drop is of a temporary nature. Moving on towards point C, the parcel gets exposed to the stellar UV field. The photodissociation of \meh{} leads -- via intermediate CH, \chh{} or \chhh{} -- to neutral C, part of which is ionised to also increase the \cp{} abundance. Finally, at point D, the photoprocesses no longer play a role, so the C and \cp{} abundances go back down. Their final values relative to $\nh$ are \scit{7}{-12} and $\sim$\ten{-14}.


\subsubsection{Nitrogen chemistry}
\label{sss:1nitrogen}
The most common nitrogen-bearing species at $t=0$ is solid \mn, with an abundance of \scit{5}{-6}. The corresponding gas-phase abundance is \scit{4}{-8}. The evolution of \mn{} parallels that of CO, because they have similar binding energies and are both very stable molecules \citep{bisschop06a}. \mn{} continues to freeze out slowly until it gets near point A in \fig{1over}, where the grain temperature of 18 K causes all \mn{} ice to evaporate. The gas-phase \mn{} remains intact along the rest of the infall trajectory and its final abundance is \scit{1}{-5}, accounting for 77\% of all nitrogen.

The second largest nitrogen reservoir at the onset of collapse is \amh{}, with solid and gas-phase abundances of \scit{8}{-6} and \scit{2}{-8}. The gas-phase abundance receives a short boost at point A due to the evaporation of \mn, followed by
\begin{align}
{\rm N}_2\ +\ {\rm He}^+\ \to\ &\ {\rm N}^+\ +\ {\rm N} +\ {\rm He}\,, \label{eq:n2+hep} \\
{\rm N}^+\ +\ {\rm H}_2\ \to\ &\ {\rm NH}^+\ +\ {\rm H}\,, \label{eq:np+h2} \\
{\rm NH}^+\ +\ {\rm H}_2^{}\ \to\ &\ {\rm NH}_2^+\ +\ {\rm H}\,, \label{eq:nhp+h2} \\
{\rm NH}_2^+\ +\ {\rm H}_2^{}\ \to\ &\ {\rm NH}_3^+\ +\ {\rm H}\,, \label{eq:nh2p+h2} \\
{\rm NH}_3^+\ +\ {\rm H}_2^{}\ \to\ &\ {\rm NH}_4^+\ +\ {\rm H}\,, \label{eq:nh3p+h2} \\
{\rm NH}_4^+\ +\ {\rm e}^-\ \to\ &\ {\rm NH}_3^{}\ +\ {\rm H}\,. \label{eq:nh4p+e}
\end{align}
The binding energy of \amh{} is intermediate to that of \mo{} and \w, so it evaporates between points B and C\@. Like \w{}, \amh{} is photodissociated upon evaporation. As the last of the \amh{} ice leaves the grains at $R=50$ AU (10 AU outside of point C), the gas-phase reservoir is no longer replenished and $x({\rm NH}_3)$ drops to $\sim$\ten{-14}. Some \amh{} is eventually reformed as the parcel gets into the disk, and the final abundance at point D is \scit{1}{-10} relative to $\nh$.

With an abundance of \scit{6}{-6}, solid NO is the third major initial nitrogen reservoir. Gaseous NO is a factor of twenty less abundant at $t=0$: \scit{3}{-7}. The NO gas is gradually destroyed prior to reaching point A by continued freeze-out and reactions with \hp{} and \hhhp. It experiences a brief gain at point A from the evaporation of OH and its subsequent reaction with N to give NO and H\@. As the parcel continues to point B, the solid NO begins to evaporate and the gas-phase abundance rises to \scit{6}{-6}. Photodissociation reactions then set in around $R=100$ AU and the NO abundance goes back down to \scit{6}{-9}. The evaporation and photodissociation of \amh{} cause a brief spike in the NO abundance through the reactions
\begin{align}
{\rm NH}_3\ +\ h\nu\ \to\ &\ {\rm NH}_2\ +\ {\rm H}\,, \label{eq:nh3+PH} \\
{\rm NH}_2\ +\ {\rm O}\ \to\ &\ {\rm HNO}\ +\ {\rm H}\,, \label{eq:nh2+o} \\
{\rm HNO}\ +\ h\nu\ \to\ &\ {\rm NO}\ +\ {\rm H}\,. \label{eq:hno+PH}
\end{align}
The evaporation of the last of the \amh{} ice at $R=50$ AU eliminates this channel and the NO gas abundance decreases to \scit{3}{-9} at point C\@. NO is now mainly sustained by the reaction between OH and N\@. As described above, the OH abundance drops sharply at $R=17$ AU, and the NO abundance follows suit. The final abundance at point D is $\sim$\ten{-14}.

The last nitrogen-bearing species from \fig{1abun} is atomic N itself. It starts at an abundance of \scit{1}{-7} and slowly freezes out to an abundance of \scit{2}{-8} just before reaching point A\@. At point A, \mn{} evaporates and is partially converted to \nnhp. The dissociative recombination of \nnhp{} mostly reforms \mn{}, but it also produces some NH and N\@. The N abundance jumps back to \scit{1}{-7} and remains nearly constant at that value until the parcel reaches point B, where NO evaporates and reacts with N to produce \mn{} and O\@. This reduces $x({\rm N})$ to a low of \scit{5}{-10} between points B and C\@. Moving in further, the parcel gets exposed to the stellar UV field, and NO and \amh{} are photodissociated to bring the N abundance to a final value of \scit{5}{-6} relative to $\nh$. As such, it accounts for 22\% of all nitrogen at the end of the collapse.


\subsection{Other parcels}
\label{subsec:other}
At the end of the collapse ($t=\tacc=\scim{2.52}{5}$ yr), the parcel from Sect.\ \ref{subsec:1parcel} (hereafter called our reference parcel) is located at $R=6.3$ AU and $z=2.4$ AU, about 0.2 AU below the surface of the disk. As shown in \fig{1over}, its trajectory passes close to the outflow wall, through a region of low extinction. This results in the photodissociation or photoionisation of many species. At the same time, the parcel experiences dust temperatures of up to 150 K (\fig{1abun}), well above the evaporation temperature of \w{} and all other non-refractory species in our network. Material that ends up in other parts of the disk encounters different physical conditions during the collapse and therefore undergoes a different chemical evolution. This subsection shows how the absence or presence of some key chemical processes, related to certain physical conditions, affects the chemical history of the entire disk. \tb{abun} lists the abundances of selected species at four points in the disk at $\tacc$.

\begin{table}
\caption{Abundances of selected species (relative to $\nh$) and physical conditions at $t=\tacc$ at four positions in the disk (two on the midplane, two at the surface).}
\label{tb:abun}
\centering
\begin{tabular}{lcccc}
\hline\hline
Species & $R= 6$ AU & $R=24$ AU & $R= 6$ AU & $R=24$ AU \\
        & $z=0$ AU   & $z=0$ AU   & $z=2$ AU   & $z=6$ AU \\
\hline
\w           & 5(-7) & 1(-12)           & 1(-8) & 4(-9) \\
\w{} ice     & 1(-4) & 1(-4) & 7(-8) & 8(-5) \\
\mo          & 1(-6) & 6(-8) & 1(-7) & 1(-5) \\
\mo{} ice    & $<$1(-12)                     & $<$1(-12)                     & $<$1(-12)                     & $<$1(-12)                     \\
O            & 1(-6) & 1(-6) & 1(-4) & 8(-7) \\
OH           & 2(-11)           & $<$1(-12)                     & $<$1(-12)                     & 9(-10)           \\
CO           & 7(-5) & 7(-5) & 8(-5) & 6(-5) \\
CO ice       & $<$1(-12)                     & $<$1(-12)                     & $<$1(-12)                     & $<$1(-12)                     \\
\meh         & 7(-8) & 1(-7) & 2(-9) & 1(-5) \\
\meh{} ice   & $<$1(-12)                     & $<$1(-12)                     & $<$1(-12)                     & $<$1(-12)                     \\
\hcop        & 1(-12)           & 5(-11)           & 3(-12)           & 4(-12)           \\
C            & $<$1(-12)                     & $<$1(-12)                     & 9(-12)           & $<$1(-12)                     \\
\cp          & $<$1(-12)                     & $<$1(-12)                     & $<$1(-12)                     & $<$1(-12)                     \\
\mn          & 1(-5) & 1(-5) & 1(-5) & 5(-6) \\
\mn{} ice    & $<$1(-12)                     & $<$1(-12)                     & $<$1(-12)                     & $<$1(-12)                     \\
\amh         & 5(-10)           & 2(-10)           & 4(-10)           & 8(-6) \\
\amh{} ice   & $<$1(-12)                     & 5(-9) & $<$1(-12)                     & 5(-8) \\
NO           & 1(-8) & 5(-10)           & $<$1(-12)                     & 6(-6) \\
NO ice       & $<$1(-12)                     & $<$1(-12)                     & $<$1(-12)                     & $<$1(-12)                     \\
\nnhp        & $<$1(-12)                     & $<$1(-12)                     & $<$1(-12)                     & $<$1(-12)                     \\
N            & 5(-11)           & 1(-9) & 5(-6) & 4(-11)           \\
\hline
$\nh$ (\pcc) & 5(12) & 8(11) & 5(11) & 4(10) \\
$\td$ (K)    & 107   & 76    & 96    & 74 \\
\w{} zone\tablefootmark{a} & 7 & 7 & 2 & 1 \\
\meh{} zone\tablefootmark{a} & 5 & 5 & 2 & 1 \\
\amh{} zone\tablefootmark{a} & 8 & 8 & 5 & 2 \\
\hline
\end{tabular}
\tablefoot{\tablefoottext{a}{The zones from Figs.\ \ref{fig:water}, \ref{fig:methane} and \ref{fig:ammonia} in which each position is located.}}
\end{table}


\subsubsection{Oxygen chemistry}
\label{sss:ooxygen}
The main oxygen reservoir at the onset of collapse is \w{} ice (Sect.\ \ref{sec:precoll}). Its abundance remains constant at \scit{1}{-4} in our reference parcel until it gets to point C in \fig{1over}, where it evaporates from the dust and is immediately photodissociated. When the parcel enters the disk, some \w{} is reformed to produce final gas-phase and solid abundances of $\sim$\ten{-8} relative to $\nh$ (Sect.\ \ref{sss:1oxygen}).

\begin{figure}
\resizebox{\hsize}{!}{\includegraphics{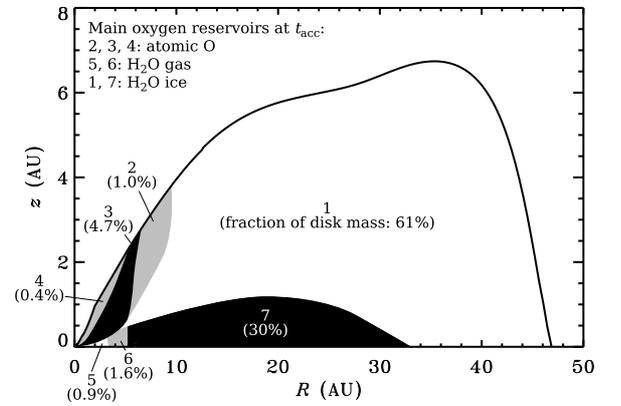}}
\caption{Schematic view of the history of \w{} gas and ice throughout the disk. The main oxygen reservoir at $\tacc$ is indicated for each zone; the histories are described in the text. The percentages indicate the fraction of the disk mass contained in each zone. Note the disproportionality of the $R$ and $z$ axes. The colours have no specific meaning other than to distinguish the different zones.}
\label{fig:water}
\end{figure}

\figg{water} shows the disk at $\tacc$, divided into seven zones according to different chemical evolutionary schemes for \w\@. The fraction of the disk mass in each zone is indicated. The material in \textbf{zone 1} (61\% of the disk mass) is the only material in the disk in which \w{} never evaporates during the collapse, because the temperature never gets high enough. The \w{} ice abundance is constant throughout zone 1 at $\tacc$ at $\sim$\scit{1}{-4}. For the material ending up in the other six zones, \w{} evaporates at some point during the collapse phase. \textbf{Zone 2} (1.0\%) contains our reference parcel, so its \w{} history has already been described. The total \w{} abundance (gas and ice combined) at $\tacc$ is $\sim$\ten{-8} at the top of zone 2 and $\sim$\ten{-6} at the bottom. The gas-ice ratio goes from $\sim$1 at the top to $\sim$\ten{-6} at the bottom.

The \w{} history of \textbf{zone 3} (4.7\%) is the same as that of zone 2, except that it finishes with a gas-ice ratio larger than unity. In both cases \w{} evaporates and is photodissociated prior to entering the disk (point C in \fig{1over}), and it partially reforms inside the disk (point D). Parcels ending up in \textbf{zone 4} (0.4\%) also experience the evaporation and photodissociation of \w{}. However, the low extinction against the stellar radiation in zone 4 prevents \w{} from reforming like it does in zones 2 and 3.

The material in \textbf{zone 5} (0.9\%) has a different history from that in zones 2--4 because it enters the disk earlier: between \scit{1.3}{5} and \scit{2.3}{5} yr. The material in zones 2--4 all accretes after \scit{2.4}{5} yr. The infall trajectories terminating in zone 5 do not pass close enough to the outflow wall or the inner disk surface for photoprocesses to play a role. All \w{} in zone 5 is in the gas phase at $\tacc$ (abundance: \scit{1}{-4}) because it lies inside the disk's snow line. The evaporation of \w{} ice does not occur until the material actually crosses the snow line. Prior to that point, the temperature never gets high enough for \w{} to leave the grains.

\begin{figure}
\resizebox{\hsize}{!}{\includegraphics{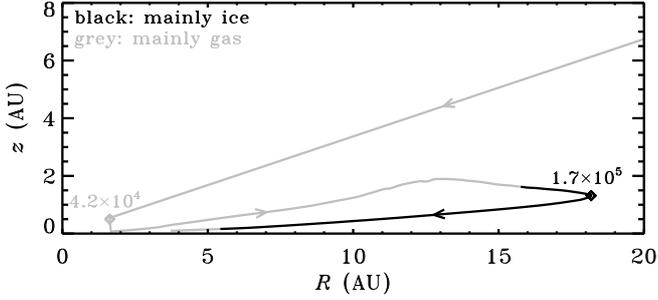}}
\caption{Infall trajectory for a parcel ending up in zone 6 from \fig{water}, showing where \w{} is predominantly present as ice (black) or gas (grey). The two diamonds mark the time in years after the onset of collapse.}
\label{fig:wtraj}
\end{figure}

The material ending up in \textbf{zones 6 and 7} (1.6 and 30\%) accretes even earlier than that ending up in zone 5: at $t=\scim{4}{4}$ yr. The disk at that time is only 2 AU large and several 100 K hot, so the ice mantles are completely removed. Upon reaching the disk, the zone 6 and 7 material is swept up as the disk expands to conserve angular momentum, resulting in trajectories like the one in \fig{wtraj}. Because the zone 5 material enters the disk at a later time, its trajectories do not feature a strong outward component and terminate at smaller radii than the zone 6 and 7 trajectories. Along the outward part of the latter, material cools down and \w{} returns to the solid phase (\fig{wtraj}). At $t=\scim{1.7}{5}$ yr, the parcel starts moving inwards again and comes close enough to the protostar for \w{} to evaporate a second time. All parcels ending up in zone 6 have similar trajectories and the same qualitative \w{} history. The parcels ending up in zone 7 also have \w{} evaporating during the initial infall and freezing out again during the outward part of the trajectory. However, they do not terminate close enough to the protostar for \w{} to desorb a second time. Therefore, most \w{} in zone 7 is on the grains at $\tacc$.

Our model is not the first one in which part of the disk contains \w{} that evaporated and readsorbed. \citet{lunine91a}, \citet{owen92a} and \citet{owen93a} argued that the accretion shock at the surface of the disk is strong enough for \w{} to evaporate. However, based on the model of \citet{neufeld94a}, \vddd{} showed that most of the disk material does not pass through a shock that heats the dust to 100 K or more. Moreover, the material that does get shock-heated to that temperature accretes close enough to the star that the stellar radiation already heats it to more than 100 K\@. Hence, including the accretion shock explicitly in our model would at most result in minor changes to the chemistry.

\begin{figure}
\resizebox{\hsize}{!}{\includegraphics{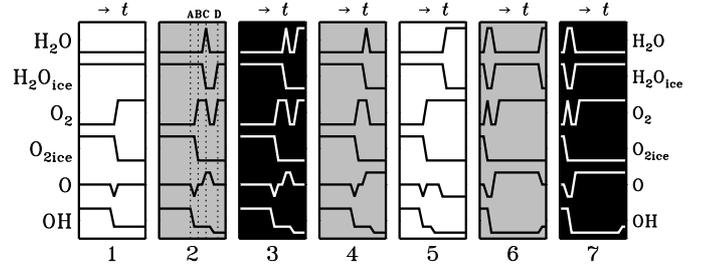}}
\caption{Qualitative evolution of some abundances towards the seven zones with different \w{} histories from \fig{water}. The horizontal axes show the time (increasing from left to right) and are non-linear. The position of points A, B, C and D from \fig{1over} is indicated for zone 2, which contains our reference parcel.}
\label{fig:scho}
\end{figure}

As discussed in Sect.\ \ref{sss:1oxygen}, \w{} controls part of the oxygen chemistry along the infall trajectory of our reference parcel, and it does so for other parcels, too. \figg{scho} presents a schematic view of the chemical evolution of six oxygen-bearing species towards each of the seven zones from \fig{water}. The abundances are indicated qualitatively as high, intermediate or low. The horizontal axes (time, increasing from left to right) are non-linear and only indicate the order in which various events take place.

For material ending up in \textbf{zone 1}, \w{} never evaporates, but \mo{} does. OH is inititally relatively abundant but most of it disappears when the overall density increases and the reactions with O, N and \hhhp{} become faster. The abundance of atomic O experiences a drop at the same time, but it quickly goes back up due to the evaporation of CO, \mo{} and NO, followed by Reactions (\ref{eq:co+hep})--(\ref{eq:o2+cn}). \mo{} also evaporates on its way to \textbf{zones 2, 3 and 4}, and because it passes through an area of low extinction, it is subsequently photodissociated. This, together with the photodissociation of OH and \w, causes an increase in the abundance of atomic O\@. Zones 2 and 3 are sufficiently shielded against the stellar UV field, so \mo{} is reformed at the end of the trajectory. This does not happen in the less extincted zone 4.

Material ending up in \textbf{zone 5} has the same qualitative history for gas-phase and solid \mo{} as has material ending up in zone 1. The evolution of atomic O initially also shows the same pattern, but it experiences a second drop as the total density becomes even higher than it does for zone 1. The higher density also causes an additional drop in the OH abundance. En route towards \textbf{zones 6 and 7}, \mo{} evaporates and is photodissociated during the early accretion onto the small disk. It is reformed during the outward part of the trajectory and survives until $\tacc$. Atomic O is also relatively abundant along most of the trajectory, although always one or two orders of magnitude below \mo. In zone 6, the O abundance decreases at the end due to the reaction with evaporating \chhh, producing \fmh{} and H\@. Zone 7 does not get warm enough for \chhh{} to evaporate, so O remains intact.


\subsubsection{Carbon chemistry}
\label{sss:ocarbon}
\begin{figure}
\resizebox{\hsize}{!}{\includegraphics{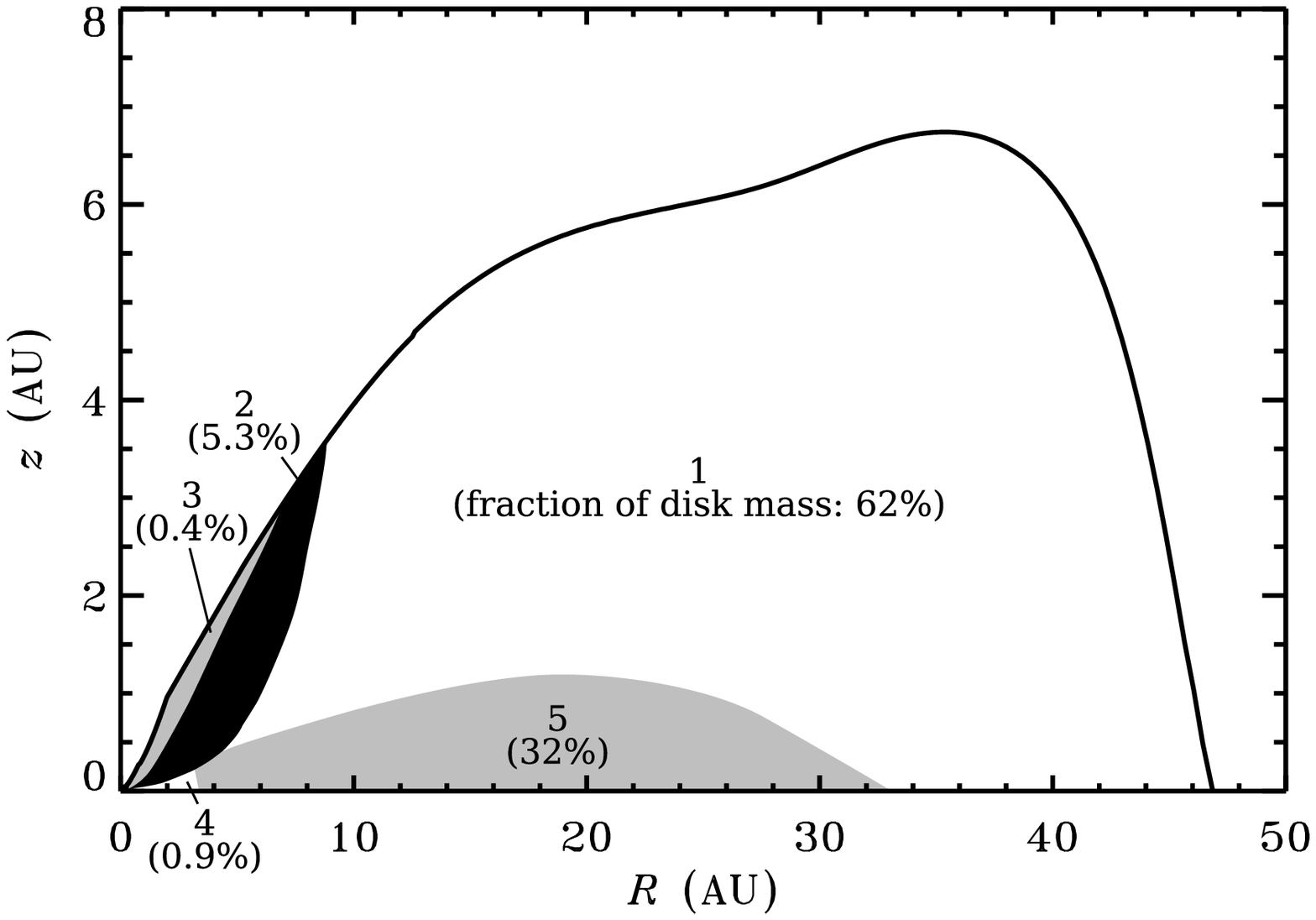}}
\caption{As \fig{water}, but for \meh{} gas and ice. The main carbon reservoir at $\tacc$ is CO gas throughout the disk.}
\label{fig:methane}
\end{figure}

The two main carbon reservoirs at the onset of collapse are CO ice and \meh{} ice (Sect.\ \ref{sec:precoll}). Their binding energies are relatively low (855 and 1080 K), so they evaporate throughout the core soon after the collapse begins (Sect.\ \ref{sss:1carbon}). For our reference parcel, the main difference between the evolution of CO and \meh{} is the photodissociation of the latter (near point C in \fig{1over}) while the former remains intact. Following the analysis for \w{}, the disk at $\tacc$ could be divided into several zones according to different chemical evolution scenarios for CO\@. However, the entire disk has the same qualitative CO history: CO does not undergo any processing apart from evaporating early on in the collapse phase, and it is the main carbon reservoir throughout ($>$99\%; \vddd). Hence, the disk is divided according to the evolution of \meh{} instead (\fig{methane}).

For material that ends up in \textbf{zone 1} (representing 62\% of the disk mass), the only major chemical event for \meh{} is the evaporation during the initial warm-up of the core. It is not photodissociated at any point during the collapse, nor does it freeze out again or react significantly with other species. Material ending up closer to the star, in \textbf{zones 2 and 3} (5.3 and 0.4\%), is sufficiently irradiated by the stellar UV field for \meh{} to be photodissociated. The amount of extinction increases towards the end of the trajectories terminating in these zones, allowing some \meh{} to reform. The final abundance ranges from \ten{-10} in zone 3 to \ten{-7} in the most shielded parts of zone 2.

\textbf{Zone 4} (0.9\%) contains material that accretes onto the disk several \ten{4} yr earlier than does the material in zones 2 and 3. It always remains well shielded from the stellar radiation, so the only processing of \meh{} is the evaporation during the early parts of the collapse. The \meh{} history of zones 1 and 4 is thus qualitatively the same. \textbf{Zone 5} (32\%) consists of material that accretes around $t=\scim{4}{4}$ yr and is subsequently transported outwards to conserve angular momentum (\fig{wtraj}). \meh{} in this material evaporates before entering the young disk and is photodissociated as it gets within a few AU of the protostar. The resulting atomic C is mostly converted into CO and remains in that form for the rest of the trajectory. Hence, even though the extinction decreases again when the parcel moves outwards, no \meh{} is reformed.

\begin{figure}
\resizebox{\hsize}{!}{\includegraphics{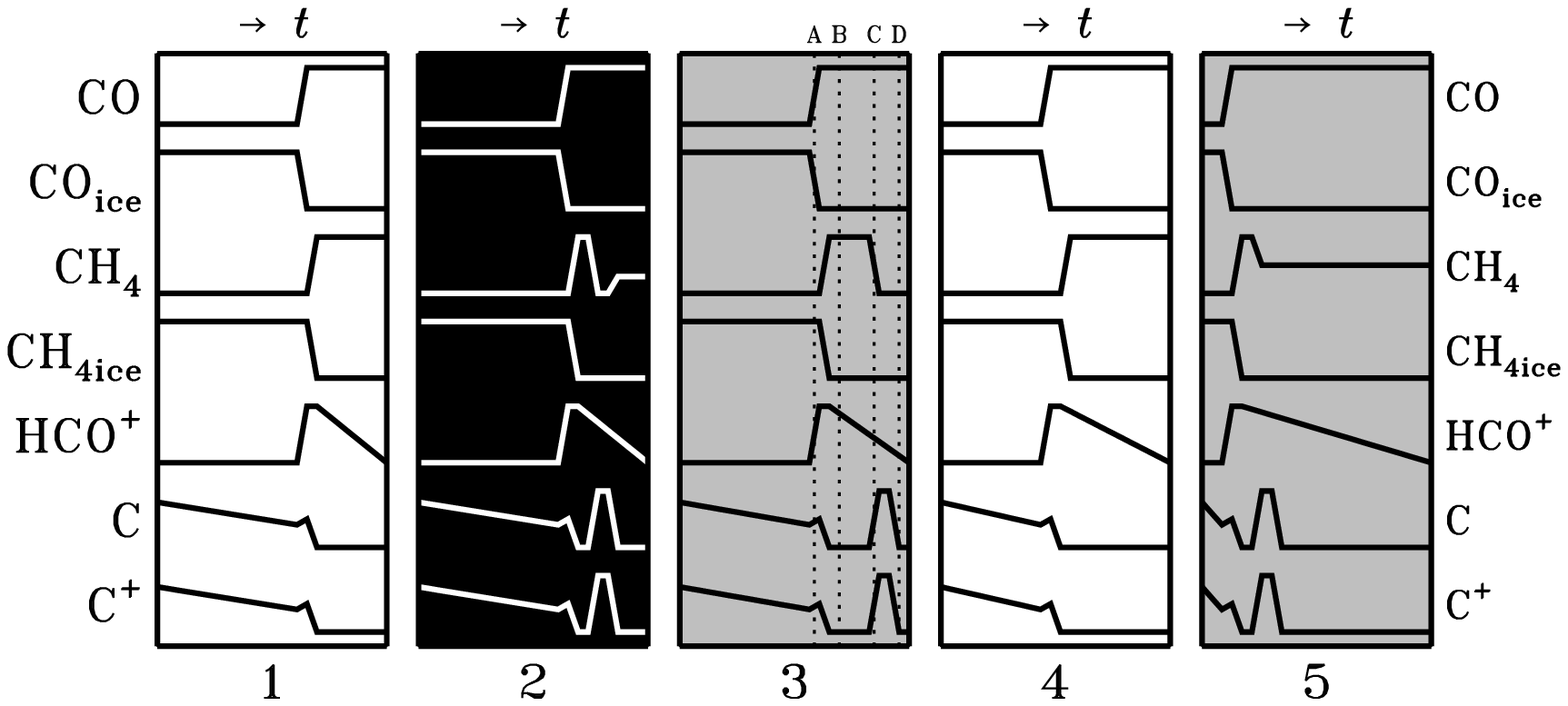}}
\caption{As \fig{scho}, but for the five zones with different \meh{} histories from \fig{methane}.}
\label{fig:schc}
\end{figure}

The evolution of the abundances of \meh{} gas and ice, CO gas and ice, \hcop, C and \cp{} towards each of the five zones is plotted schematically in \fig{schc}. The \hcop{} abundance follows the CO abundance at a ratio that is roughly inversely proportional to the overall density. Hence, the \hcop{} evolution is qualitatively the same towards each zone: it reaches a maximum abundance of a few \ten{-10} when CO evaporates and gradually disappears as the density increases along the rest of the infall trajectories. The most complex history amongst these seven carbon-bearing species is found in C and \cp. Towards all five zones, they are initially destroyed by reactions with \mh, \mo{} and OH\@. Some C and \cp{} is reformed when CO evaporates, but the subsequent evaporation of \mo{} and OH causes the abundances to decrease again. En route to zones 2, 3 and 5, the photodissociation of \meh{} leads to a second increase in C and \cp{}, followed by a third and final decrease when the parcel moves into a more shielded area.


\subsubsection{Nitrogen chemistry}
\label{sss:onitrogen}
Most nitrogen at the onset of collapse is in the form of solid \mn{} (41\%), solid \amh{} (32\%) and solid NO (22\%). The evolution of \mn{} during the collapse is the same as that of CO, except for a minor difference in the binding energy. Both species evaporate shortly after the collapse begins and remain in the gas phase throughout the rest of the simulation. Neither one is photodissociated, because the protostar is not hot enough to provide sufficient UV flux shortwards of 1100 \AA.

\begin{figure}
\resizebox{\hsize}{!}{\includegraphics{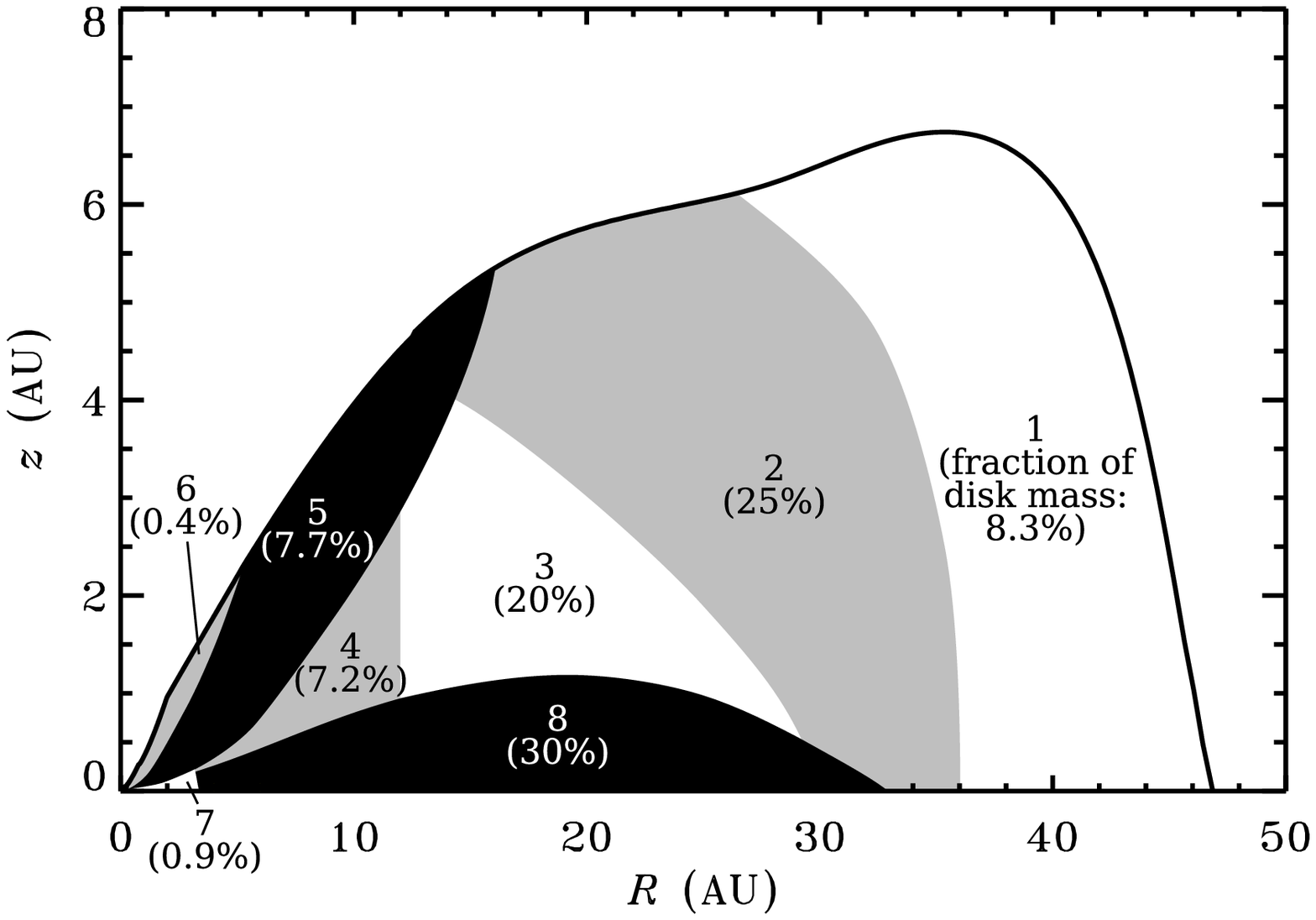}}
\caption{As \fig{water}, but for \amh{} gas and ice. The main nitrogen reservoir at $\tacc$ is \mn{} gas throughout the disk.}
\label{fig:ammonia}
\end{figure}

The evolution of \amh{} shows a lot more variation, as illustrated in \fig{ammonia}. The disk at $\tacc$ is divided into eight zones with different \amh{} histories. No processing occurs towards \textbf{zone 1} (8.3\% of the disk mass): the temperature never exceeds the 73 K required for \amh{} to evaporate, so it simply remains on the grains the whole time. Material ending up in \textbf{zone 2} (25\%) does get heated above 73 K, and \amh{} evaporates. However, it freezes out again at the end of the trajectory because zone 2 itself is not warm enough to sustain gaseous \amh. The final solid \amh{} abundance in zones 1 and 2 is about \scit{8}{-6} relative to $\nh$ (\tb{abun}).

\amh{} ending up in \textbf{zone 3} (20\%) evaporates from the grains just before entering the disk, but it is immediately photodissociated or destroyed by \hcop. Towards the end of the trajectory, some \amh{} is reformed from the dissociative recombination of \amhhp. It rapidly freezes out to produce a final solid \amh{} abundance of \ten{-10}--\ten{-8}. \textbf{Zone 4} (7.2\%) has the same history, except that there is an additional adsorption-desorption cycle before the destruction by photons and \hcop.

Our standard parcel ends up in \textbf{zone 5} (7.7\%); its \amh{} evolution is mainly characterised by photodissociation above the disk and reformation inside it (Sect.\ \ref{sss:1nitrogen}). \amh{} evaporates when it gets to within about 200 AU of the star, halfway between points B and C in \fig{1over}. It is immediately photodissociated, but some \amh{} is reformed once the material is shielded from the stellar radiation. The reformed \amh{} remains in the gas phase. No reformation takes place in the less extincted \textbf{zone 6} (0.4\%), which otherwise has the same \amh{} history as zone 5.

Material ending up in \textbf{zone 7} (0.9\%) does not pass close enough to the outflow wall for \amh{} to be photodissociated upon evaporation. Instead, \amh{} is mainly destroyed by \hcop{} and attains a final abundance of $\sim$\ten{-9}. Lastly, \textbf{zone 8} (30\%) contains again the material that accretes onto the disk at an early time and then moves outwards to conserve angular momentum. Its \amh{} evaporates already before reaching the disk and is subequently dissociated by the stellar UV field. As the material moves away from the star and is shielded from its radiation, some \amh{} is reformed out of \amhhp{} to a final abundance of $\sim$\ten{-10}.

\begin{figure}
\resizebox{\hsize}{!}{\includegraphics{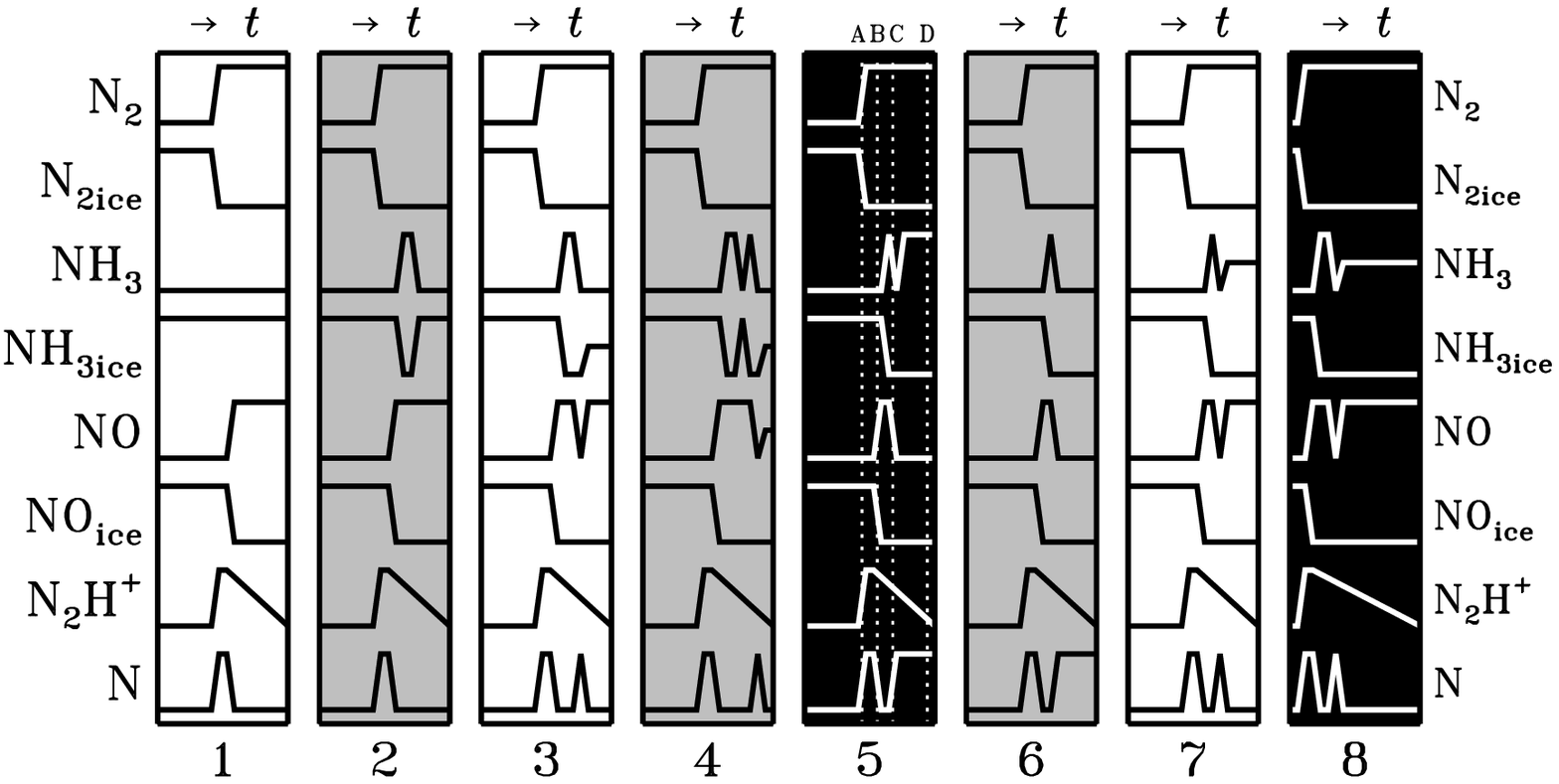}}
\caption{As \fig{scho}, but for the eight zones with different \amh{} histories from \fig{ammonia}.}
\label{fig:schn}
\end{figure}

The abundances of \nnhp{} and atomic N are largely controlled by the evolution of \mn{} and \amh{}, as shown schematically in \fig{schn}. In all parcels, regardless of where they end up, \nnhp{} is mainly formed out of \mn{} and \hhhp, so its abundance goes up when \mn{} evaporates shortly after the onset of collapse. It gradually disappears again as the collapse proceeds due to the increasing density. The atomic N abundance at the onset of collapse is \scit{1}{-7}. It increases when \mn{} evaporates and decreases again a short while later when NO evaporates (Sect.\ \ref{sss:1nitrogen}). For material that ends up in zones 1 and 2, the N abundance is mostly constant for the rest of the collapse at a value of \ten{-13} (inner part of zone 2) to \ten{-10} (outer part of zone 1). Material that ends up in the other six zones is exposed to enough UV radiation for NO and \amh{} to be photodissociated, so there is a second increase in atomic N\@. Zones 3, 4, 7 and 8 are sufficiently shielded at $\tacc$ to reform some or all NO and \amh{}, and the N abundance finishes low. Less reformation is possible in zones 5 and 6, so they have a relatively large amount of atomic N at the end of the collapse phase.


\subsubsection{Mixing}
\label{sss:mix}
Given the dynamic nature of circumstellar disks, the zonal distribution presented in the preceding subsections may offer too simple a picture of the chemical composition. For example, there are as yet no first principles calculations of the processes responsible for the viscous transport in disks. The radial velocity equation used in our model \citep{visser10a} is suitable as a zeroth-order description, but it omits the effects of mixing \citep{lewis80a,semenov06a,aikawa07a} and it cannot explain important observational features like episodic accretion \citep{kenyon95a}. Indeed, with the \citet{toomre64a} $Q$ parameter reaching a minimum of 1.4 at $\tacc$, the disk is only barely gravitationally stable. Small changes in the model or in the initial conditions would suffice to make the disk unstable at various points during the collapse phase, resulting in a redistribution of material beyond the equations currently used. Hence, in real disks, both the shapes and the locations of the zones are likely to differ from what is shown in Figs.\ \ref{fig:water}, \ref{fig:methane} and \ref{fig:ammonia}. A larger degree of mixing would also make the borders between the zones more diffuse than they are in our simple schematic representation. Nevertheless, the general picture from this section offers a plausible description of the chemical history towards different parts of the disk. Spectroscopic observations at 5--10 AU resolution, as possible with ALMA, are required to determine to what extent this picture holds in reality. Based on the abundances in \tb{abun}, potential tracers are \w, NO and \amh.


\section{Chemical history versus local chemistry}
\label{sec:local}
Section \ref{sec:coll} contains many examples of abundances increasing or decreasing on short timescales of less than a hundred years (see, e.g., \fig{1abun}). It appears that the abundances respond rapidly to the changing physical conditions as material falls in supersonically through the inner envelope and accretes onto the disk. However, this does not necessarily mean that the abundances are always in equilibrium. The current section explores the question of whether the disk is in chemical equilibrium at the end of the collapse, or whether its chemical composition is a non-equilibrium solution to the conditions encountered during the collapse. To that end, the chemistry is evolved for an additional 1 Myr beyond $\tacc$. The density, temperature, UV flux and extinction are kept constant at the values they have at $\tacc$, and all parcels of material are kept at the same position. Clearly, this is a purely hypothetical scenario. In reality, the disk would change in many ways after $\tacc$: it spreads in size, the surface layers become more strongly irradiated, the temperature changes, the dust coagulates into planetesimals, gas is photoevaporated from the surface layers, and so on. All of these processes have the potential to affect the chemical composition. However, they would also interfere with the attempt to determine whether the disk is in chemical equilibrium at $\tacc$. This question is most easily addressed by evolving the chemistry for an additional period of time at constant conditions. Hence, all of the physical changes that would occur during the post-collapse phase are ignored.

\begin{figure}[t!]
\resizebox{\hsize}{!}{\includegraphics{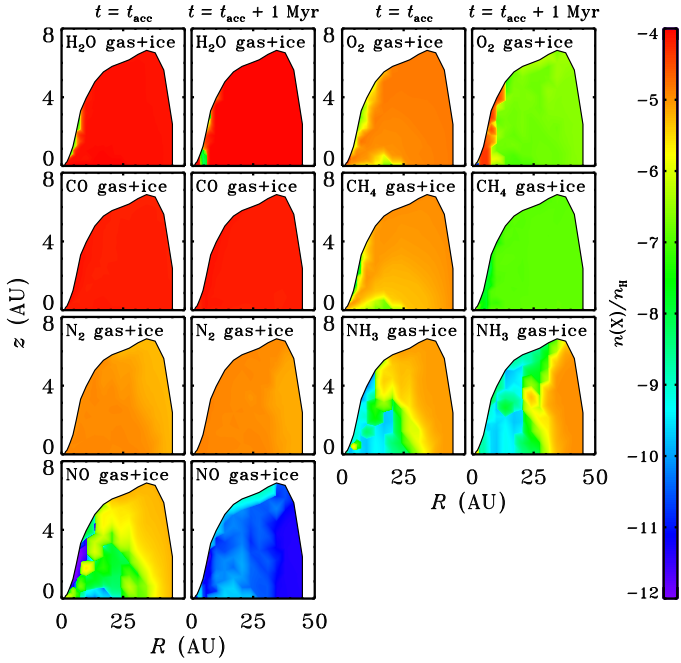}}
\caption{Abundances of total \w, \mo, CO, \meh, \mn, \amh{} and NO (gas and ice combined) throughout the disk at the end of the collapse phase ($t=\tacc$) and after an additional 1 Myr post-collapse phase ($t=\tacc+1$ Myr).}
\label{fig:cvsgi}
\end{figure}

The 21 species discussed in Sect.\ \ref{sec:coll} can be divided into two categories: those whose abundance profile changes during the post-collapse phase, and those whose abundance profile remains practically the same. Members of the ``changed'' category are \mo, OH, \meh, \amh{} and NO (all gaseous), \amh{} ice, and atomic O and N\@. The thirteen species in the ``unchanged'' category are \w{} gas and ice, \mo{} ice, CO gas and ice, \meh{} ice, C, \cp, \mn{} gas and ice, NO ice, \hcop{} and \nnhp. The individual gas and ice abundances are summed in \fig{cvsgi} and compared at $t=\tacc$ and $\tacc+1$ Myr. The total \w, CO and \mn{} abundances do not change significantly during the post-collapse phase, while the total \mo, \meh, \amh{} and NO abundances change by more than two orders of magnitude in a large part of the disk.

There are two areas in the disk where the abundances of all 21 species remain nearly constant: near the surface out to $\sim$10 AU, and at the midplane between $\sim$5 and $\sim$25 AU\@. The chemistry in the first area is dominated by fast photoprocesses. The second area, near the midplane, is the densest part of the disk and therefore has high collision frequencies. In both cases, the chemical timescales are short and the chemistry reaches equilibrium during the final stages of accretion.

Outside these two ``equilibrium areas'', one of the key processes during the post-collapse phase is the conversion of gas-phase oxygen-bearing species into \w{}, which subsequently freezes out. Because \w{} ice is already one of the main oxygen reservoirs at $\tacc$ throughout most of the disk (\fig{water}), its abundance only increases by about 20\% during the rest of the simulation. These 20\% represent 10\% of the total oxygen budget and correspond to \scit{2}{-5} oxygen nuclei per hydrogen nucleus.

One of the species converted into \w{} ice is NO, which reacts with \nhh{} through
\begin{equation}
\label{eq:nh2+no}
{\rm NH}_2\ +\ {\rm NO}\ \to\ {\rm N}_2\ +\ {\rm H}_2{\rm O}\,,
\end{equation}
followed by adsorption of \w{} onto dust. This reaction is responsible for the post-collapse destruction of gas-phase NO outside the two ``equilibrium areas'', as well as for a $\sim$20\% increase in gas-phase \mn. NO is the main destructor of atomic N via
\begin{equation}
\label{eq:no+n-3}
{\rm NO}\ +\ {\rm N}\ \to\ {\rm N}_2\ +\ {\rm O}\,,
\end{equation}
so the conversion of NO into \w{} leads to a higher N abundance. The \nhh{} required for \rx{nh2+no} is formed from \amh:
\begin{align}
{\rm NH}_3^{}\ +\ {\rm HCO}^+\ \to\ &\ {\rm NH}_4^+\ +\ {\rm CO}\,, \label{eq:nh3+hcop} \\
{\rm NH}_4^+\ +\ {\rm e}^-\ \to\ &\ {\rm NH}_2^{}\ +\ {\rm H}_2^{}/2{\rm H}\,. \label{eq:nh4p+e-2}
\end{align}
The gas-phase reservoir of \amh{} is continously fed by evaporation of \amh{} ice, so Reactions (\ref{eq:nh3+hcop}), (\ref{eq:nh4p+e-2}) and (\ref{eq:nh2+no}) effectively transform all \amh{} into \mn{}.

Two other important species that are destroyed in the post-collapse phase are gaseous \mo{} and \meh. \mo{} is ionised by cosmic rays to produce \oop, which reacts with \meh:
\begin{align}
{\rm CH}_4^{}\ +\ {\rm O}_2^+\ \to\ &\ {\rm HCOOH}_2^+\ +\ {\rm H}\,, \label{eq:ch4+o2p} \\
{\rm HCOOH}_2^+\ +\ {\rm e}^-\ \to\ &\ {\rm HCOOH}\ +\ {\rm H}\,. \label{eq:hcooh2p+e}
\end{align}
The formic acid (HCOOH) freezes out, thus acting as a sink for both carbon and oxygen. At $t=\tacc+1$ Myr, the solid HCOOH abundance is \scit{3}{-6}, accounting for 3\% of all oxygen and 4\% of all carbon. The presence of such a sink is a common feature of disk chemistry models \citep[e.g.,][]{aikawa99a}. Gas-phase species are processed by He$^+$ and \hhhp{} until they form a species whose evaporation temperature is higher than the dust temperature. This is HCOOH in our case, but it could also be carbon-chain molecules like C$_2$H$_2$ or C$_3$H$_4$.

Another important post-collapse sink reaction in our model is the freeze-out of HNO\@. At $t>\tacc$, OH is primarily formed by
\begin{equation}
\label{eq:hno+o}
{\rm HNO}\ +\ {\rm O}\ \to\ {\rm NO}\ +\ {\rm OH}
\end{equation}
throughout most of the disk. This reaction is also one of the main destruction channels for atomic O\@. Hence, the gradual freeze-out of HNO after $\tacc$ leads to a decrease in OH and an increase in O\@. Also contributing to the higher O abundance is the fact that OH itself is an important destructor of O through \rx{oh+o}.

The post-collapse chemistry described here is merely meant as an illustration of what might take place in a real circumstellar disk after the main accretion phase comes to an end. Depending on how the disk continues to evolve physically, other reactions may become more important than the ones listed above. However, with the abundances of several key species changing by orders of magnitude in our post-collapse experiment, the disk was clearly not in chemical equilibrium at the end of the collapse. Even the two ``equilibrium areas'' (near the surface out to 10 AU and from 5 to 25 AU at the midplane) are not truly in chemical equilibrium when considering less abundant species. For example, at 20 AU on the midplane, post-collapse processing increases the abundance of solid HCOOH by a factor of 13 (from \scit{7}{-8} to \scit{9}{-7} relative to $\nh$) and that of solid \meoh{} by a factor of 27 (from \scit{3}{-13} to \scit{8}{-12}).

The lack of chemical equilibrium at the end of the collapse phase means that one has to be careful in choosing the initial abundances when running T Tauri or Herbig Ae/Be disk chemistry models. The abundances of minor species, such as most organic molecules, can be particularly sensitive to what happened before the disk was formed. Ideally, any disk chemistry model should therefore include a realistic core collapse phase to set the stage for the chemical evolution in the disk itself.


\section{Comets: mixed origins?}
\label{sec:comets}
\begin{figure}[t!]
\resizebox{\hsize}{!}{\includegraphics{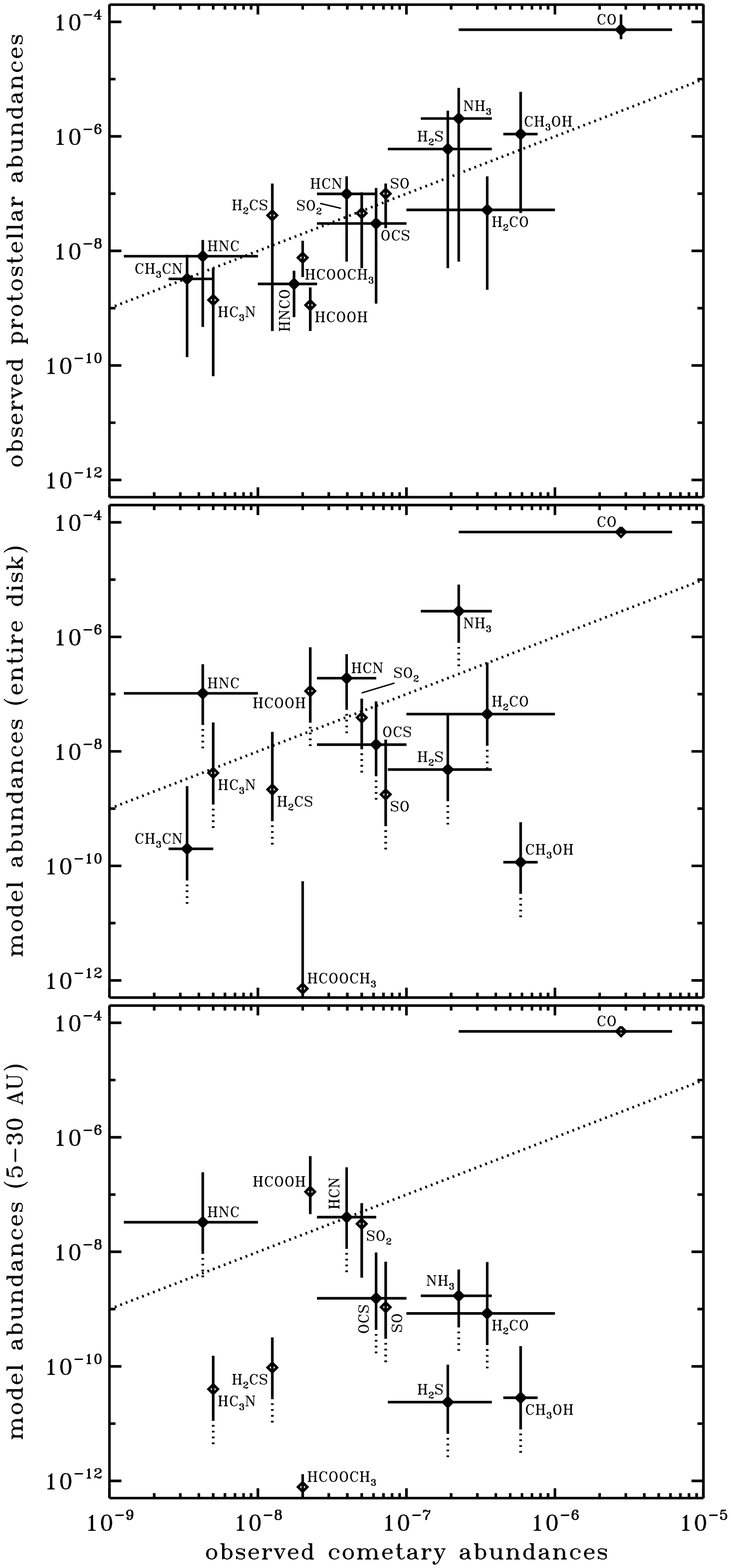}}
\caption{\emph{Top:} abundances observed in comets versus gas-phase abundances observed in young protostars. \emph{Middle and bottom:} abundances observed in comets versus total abundances (gas and ice) from our model at $t=\tacc$. All abundances are relative to $\nh$. For the observational data points, the error bars indicate the spread between sources. For the model data, they indicate the range of values across the entire disk (\emph{middle}) or the comet-forming region (\emph{bottom}). Dotted error bars indicate the full range extends to less than \ten{-12}. The diagonal dotted lines represent hypothetical one-to-one relationships between abundances on the $x$ and $y$ axes.}
\label{fig:cvi}
\end{figure}

The chemical composition of cometary ices shows many similarities to that of young protostars, but the abundance of a given species may vary by more than an order of magnitude from one comet to the next \citep{ahearn95a,mumma11a}. Both points are illustrated in the top panel of \fig{cvi}, where the abundances of several species in the comets \object{1P/Halley}, \object{C/1995 O1} (Hale-Bopp), \object{C/1996 B2} (Hyakutake), \object{C/1999 H1} (Lee), \object{C/1999 S4} (LINEAR) and \object{153P/Ikeya-Zhang} are plotted against the gas-phase abundances in the embedded protostar \object{IRAS 16293--2422} (warm inner envelope), the bipolar outflow \object{L~1157}, and the four hot cores \object{W3(\w)}, \object{G34.3+0.15}, \object{Orion HC} and \object{Orion CR}\@. All abundances are gathered from \citet{bockelee00a,bockelee04a} and \citet{schoier02a}. The cometary data are converted from abundances relative to \w{} to abundances relative to $\nh$ by assuming $n(\wm)/\nh=\scim{5}{-5}$.

Each point in the top panel of \fig{cvi} is characterised by the mean value from the available sources (the diamonds) and the total spread in measurements (the error bars). Uncertainties from individual measurements are not shown. The dotted line marks the one-to-one relationship where the cometary abundances equal the protostellar abundances. The data generally follow this line, suggesting that the material ending up in comets underwent little chemical processing during the evolution from protostar to main-sequence star. However, the plot also shows variations of at least an order of magnitude in the cometary abundances for CO, \fmh, \meoh, HNC and \hhs, as well as smaller variations for other species. A possible explanation for the different chemical compositions is that the comets formed in different parts of the solar nebula. If that is indeed the case, there must have been some degree of chemical processing between the protostellar stage and the time where each comet formed. Another point to note is that the elemental nitrogen abundance in comets is at least a factor of three lower than it is in protostars \citep{wyckoff91a}. Although the reason for this deficiency remains unclear, it also points at a certain degree of chemical processing.

Comets are thought to have formed at the gravitational midplane of the circumsolar disk, between 5 and 30 AU from the young Sun \citep{bockelee04a}. In our model disk, the 5--30 AU range is almost exactly the area containing material that accreted at an early time (\scit{4}{4} yr after the onset of collapse) and stayed in the disk for the remainder of the collapse phase (\fig{wtraj}). As discussed in Sect.\ \ref{subsec:other}, this material undergoes a larger degree of chemical processing than any other material in the disk, due to its accreting close to the star. However, such a large degree of processing is hard to reconcile with the observed similarities between cometary and protostellar abundances, and it is therefore unlikely that comets form entirely out of material that accreted close to the star. Instead, as argued below, our model results suggest that comets consist of a mix of processed and unprocessed material.

The middle and bottom panels of \fig{cvi} compare the cometary abundances from \citet{bockelee00a,bockelee04a} to the volume-averaged abundances from our model at the end of the collapse phase. For all species, the gas and ice abundances from the model are summed and displayed as one. The horizontal error bars show again the spread in abundances between individual comets. In the middle panel, the vertical bars show the spread across the entire model disk; in the bottom panel, they show the spread in the 5--30 AU region at the midplane. Dotted extensions to the error bars indicate that the range of values extends to less than \ten{-12}.

The disk-averaged abundances tend to cluster around the hypothetical one-to-one relationship indicated by the diagonal dotted line in the middle panel of \fig{cvi}. Two notable exceptions are \meoh{} and \mf. Both species are predominantly formed through grain-surface chemistry, which is not included in our model. The spread in model abundances is at least six orders of magnitude, but given the wide range of physical conditions that are sampled, such a large spread is to be expected. The 5--30 AU abundances from the bottom panel of \fig{cvi} are lower for many species than the disk-wide abundances. For example, the average \amh{} abundance decreases by more than three orders of magnitude. \amh{} is photodissociated along the trajectories that terminate in the comet-forming zone, whereas other trajectories are more shielded (Sect.\ \ref{sss:onitrogen}).

The abundances in the bottom panel of \fig{cvi} are sampled over a smaller region, so they show less variation than those in the middle panel. However, they still take on a wider range of values than do the cometary abundances. More importantly, the 5--30 AU data show no correlation with the comet data. There is no indication from Sect.\ \ref{sec:local} that post-collapse processing brings the abundances back to near what they were in the protostellar stage. Hence, the midplane material between 5 and 30 AU in our model does not appear to be analogous to the material from which the solar-system comets formed.

How plausible is the large degree of chemical processing for material that ends up in the comet-forming zone? The amount of processing depends on the range of physical conditions encountered along a given infall trajectory, which in turn depends mostly on how close to the protostar the trajectory gets. The physics in our model are kept as simple as possible, so a trajectory similar to the one drawn in \fig{wtraj} may not be fully realistic. On the other hand, the back-and-forth motion results from the well-known concept of conservation of angular momentum. As the inner parts of the disk accrete onto the star, the outer parts must move out to maintain the overall angular momentum. They may be pushed inwards again at a later time if a sufficient amount of mass is accreted from the envelope at larger radii. This happens in our simple model, but also in the more physically realistic hydrodynamical simulations of \citet{brinch08a}.

Accepting some degree of back-and-forth motion as a common part in disk evolution, the question remains how close to the star the material gets before moving out to colder parts of the disk. Our revised solution to the problem of sub-Keplerian accretion \citep{visser10a} results in radial velocity profiles that are different from the ones in \vddd\@. With the old profiles, all material that accretes inside of the snow line always remains inside of the snow line. The comet-forming zone therefore did not contain any material in which \w{} had evaporated and readsorbed onto the grains. Such material does exist in the comet-forming zone with the new velocity profiles (\fig{wtraj}). However, the result depends strongly on the initial physical conditions. If the new model is run for the parameter grid from \vddd, there are three cases (out of eight) where material accretes inside of the snow line and then moves out beyond it; the standard disk discussed throughout this work is one of them. The critical question is whether the point at which the radial velocity in the disk changes sign (from inwards to outwards) lies inside or outside of the snowline. In general, the radial velocity changes sign at smaller radii for less massive disks. Such systems are therefore more likely to have material accreting inside of the snowline and then moving out. Indeed, the three cases in our parameter grid where this happens are the three with the least massive disks. Quantitatively, the dividing factor appears to be a ratio of 0.2 between the disk mass at $t=\tacc$ and the core mass at $t=0$. For larger ratios, material that accretes inside of the snow line always remains there.

The presence of crystalline silicate dust in disks provides a strong argument in favour of trajectories similar to the one from \fig{wtraj}. Observations show crystalline fractions at $R\approx10$ AU that are significantly larger than what is found in the ISM \citep{bouwman08a}. Crystallisation of amorphous silicates requires temperatures of 800 K. The disk is less than 100 K at 10 AU, so a possible explanation for the high crystalline fractions is that material accreted in the much hotter inner disk and then moved outwards. Although the particular trajectory from \fig{wtraj} does not reach the required temperature of 800 K, other trajectories do. Crystalline dust has also been detected in several comets, including C/1995 01 (Hale-Bopp) and 81P/Wild 2, providing direct evidence that part of their constituent material has been heated to temperatures well above the evaporation temperature of \w{} \citep{wooden99a,keller06a}.

If Hale-Bopp, Wild 2 and other comets contain crystalline silicates, they must also contain ices that underwent a large degree of chemical processing. Likewise, the presence of amorphous silicates in comets is indicative of chemically unprocessed ices. Hence, the material from which comets are formed must be of mixed origins: some of it accreted close to the star and was heated to high temperatures, while another part accreted at larger radii and remained cold. Within the context of our model, this requires that material from beyond $R=30$ AU is radially mixed into the comet-forming zone between 5 and 30 AU\@. Alternatively, vertical mixing may bring relatively pristine material from higher up in the disk down into the comet-forming zone at the midplane. As for the chemical variations between individual comets, our model shows many examples of abundances changing with position or with time. The variations can thus be explained by having the comets form at different positions in the circumsolar disk, or at different times during the disk's lifetime.


\section{Caveats}
\label{sec:cav}
One of the model's three important caveats is that it sets the gas temperature equal to the dust temperature, rather than calculating it explicitly. This approach is valid for the interior of the disk (optically thick to UV radiation), but it breaks down in the surface layers and in the inner parts of the envelope \citep{kamp04a}. The chemistry in these optically thin regions is mostly controlled by photoprocesses and ion-molecule reactions, neither of which depend strongly on temperature. One aspect that would be affected is the gas-phase production of \w{} through the reactions
\begin{align}
{\rm O}\ +\ {\rm H}_2\ \to\ &\ {\rm OH}\ +\ {\rm H}\,, \label{eq:o+h2} \\
{\rm OH}\ +\ {\rm H}_2\ \to\ &\ {\rm H}_2{\rm O}\ +\ {\rm H}\,, \label{eq:oh+h2}
\end{align}
which have activation barriers of 3100 and 2100 K, respectively \citep{natarajan87a,atkinson04a}. Hence, our model probably underestimates the amount of \w{} in the surface of the disk.

A higher gas temperature would also change some of the physics in the model, such as increasing the pressure of the disk. This would push up the disk-envelope boundary and possibly change the spatial distribution of where material accretes onto the disk. However, the bulk of the accretion currently takes place in shielded regions, so only a small fraction of the infalling material is expected to be affected.

Another caveat is the adopted shape of the stellar spectrum, which may have larger chemical consequences than the gas temperature. Presently, the stellar radiation field is a blackbody spectrum at the effective stellar temperature. Our star never gets hotter than 5800 K (\vddd), which is not enough to produce UV photons of sufficient energy to dissociate CO and \mh. Many T Tauri stars have a UV excess \citep{herbig86a}, which \emph{would} allow CO and \mh{} to be photodissociated. The higher UV flux would also enhance the photoionisation of atomic C, probably resulting in \cp{} being the dominant form of carbon between points C and D in \fig{1over}. In turn, this would boost the production of carbon-chain species like C$_2$S and HC$_3$N\@. As soon as the material enters the disk and is shielded again from the stellar radiation, \cp{} is converted back into C and CO\@. However, some signature of the temporary high \cp{} abundance may survive.

The third important caveat has to do with the dust acting as a possible sink for the gas-phase chemistry. In our model, once an atom or molecule freezes out, there are three possibilities: (1) it may stay on the grain if the temperature and incident UV flux remain low; (2) it may desorb again if the temperature or UV flux gets high enough; or (3) it may be hydrogenated in a few select cases (Sect.\ \ref{subsec:grains}). In reality, there are at least three additional possibilities: (4) volatile species like CO, \meh{} and \mn{} can be trapped in the \w{} ice matrix; (5) the atom or molecule in question may undergo grain-surface reactions not included in our model; or (6) the dust grain may grow and settle to the midplane, effectively removing the molecules in its icy mantle from the active chemistry elsewhere. The latter scenario has been invoked to explain the low \w{} abundance in the circumstellar disk of \object{DM Tau} \citep{bergin10a}. Processes 4, 5 and 6 can all act as a sink of gas-phase CO and other abundant species, potentially diminishing their dominant role in the gas phase. A quantitative treatment of each process is required to assess how much they would alter the chemical evolution presented here.


\section{Conclusions}
\label{sec:conc}
This paper describes the two-dimensional chemical evolution during the collapse of a molecular cloud core to form a low-mass protostar and a circumstellar disk. The model, described in detail by \citet{visser09a} and \citet{visser10a}, consists of semi-analytical density and velocity profiles throughout the core and the disk. The dust temperature and the UV radiation field -- both important for the chemical evolution -- are calculated with a full radiative transfer method. The chemical part of the model features a full gas-phase network, including adsorption and desorption from dust grains, as well as basic hydrogenation reactions on the grain surfaces. Starting from realistic initial conditions, the chemical evolution is computed for parcels of material terminating at selected positions in the disk. Special attention is paid to parcels ending up in the comet-forming zone. Our conclusions are as follows:

\begin{itemize}
\item[$\bullet$] The chemistry during the collapse phase is controlled by a small number of key chemical processes, each of which is activated by changes in the physical conditions. The evaporation of CO, \meh{} and \w{} at approximately 18, 22 and 100 K is one set of such key processes. Another set is the photodissociation of \meh{} and \w{} (Sect.\ \ref{subsec:1parcel}).

\item[$\bullet$] At the end of the collapse phase, the disk can be divided into several zones with different chemical histories. The different histories are related to the presence or absence of the aforementioned key processes along various infall trajectories. Spectroscopic observations at high spatial resolution are required to determine whether this zonal division really exists, or if it is smoothed out by mixing (Sect.\ \ref{subsec:other}).

\item[$\bullet$] Part of the material that accretes onto the disk at early times is transported outwards to conserve angular momentum, and may remain in the disk for the rest of the collapse phase. It is heated to well above 100 K as it accretes close to the star, so \w{} and all other non-refractory species evaporate from the grains. They freeze out again when the material cools down during the subsequent outward transport (Sect.\ \ref{subsec:other}).

\item[$\bullet$] When the chemistry is evolved for an additional 1 Myr at fixed physical conditions after the end of the collapse phase, the abundances of most species change throughout the disk. Hence, the disk is not in chemical equilibrium at the end of the collapse, and one should be careful in choosing the initial abundances when running stand-alone disk chemistry models (Sect.\ \ref{sec:local}).

\item[$\bullet$] Material that ends up in the comet-forming zone undergoes a large degree of chemical processing, including the evaporation and readsorption of \w{} and species trapped in the \w{} ice. This is consistent with the presence of crystalline silicates in comets. However, it is inconsistent with the chemical similarities observed between comets and young protostars, which are indicative of little processing. Hence, it appears that the cometary material is of mixed origins: part of it was strongly processed, and part of it was not. The chemical variations observed between individual comets suggest they were formed at different positions or times in the solar nebula. Fully pristine ices only appear in the upper and outer parts of this particular disk model, although turbulent mixing may move them into the comet-forming zone (Sect.\ \ref{sec:comets}).
\end{itemize}


\begin{acknowledgements}
Astrochemistry in Leiden is supported by a Spinoza Grant from the Netherlands Organization for Scientific Research (NWO) and a NOVA grant. RV acknowledges partial support from NASA through RSA award No.\ 1371476 and from the NSF through grant AST-1008800. SDD acknowledges support by NASA grant NNX08AH28G.
\end{acknowledgements}


\bibliographystyle{aa}
\bibliography{collgas}

\end{document}